\documentclass[aps,prc,twocolumn,floatfix]{revtex4}
\usepackage{epsfig}

\begin{document}

\title{J/$\psi$ production in proton induced collisions at FAIR}
\author{Partha Pratim Bhaduri}
\affiliation{Variable Energy Cyclotron 
Centre, 1/AF Bidhan Nagar, Kolkata 700 064, India}            
\author{A. K. Chaudhuri}
\affiliation{Variable Energy Cyclotron 
Centre, 1/AF Bidhan Nagar, Kolkata 700 064, India} 
\author{Subhasis Chattopadhyay}
\affiliation{Variable Energy Cyclotron 
Centre, 1/AF Bidhan Nagar, Kolkata 700 064, India}            
\date{\today}
\begin{abstract}

We have examined the production of J/$\psi$ mesons in high energy proton-proton and proton-nucleus collisions at beam energies in the range from 158 GeV to 920 GeV, available from different fixed target experiments. In the employed model J/$\psi$ production in hadronic collisions is assumed to be a factorisable two step process: (i) production of a $c\bar{c}$ pair which can be reliably described by perturbative QCD, and  (ii) formation of J/$\psi$ resonance from the $c\bar{c}$ pair, which can be conveniently parameterized incorporating different existing physical mechanisms of color neutralization. We show that, for lower collision energies, J/$\psi$ production through quark-anti-quark annihilation gives larger contribution at higher $x_F$, while gluon-gluon fusion dominates the production at smaller $x_F$. For proton-nucleus collisions the model takes into account both the initial state modification of parton distributions in nuclei and the final state interaction of the produced $c\bar{c}$ pairs with the target nucleons. The model is found to give reasonable description of data on J/$\psi$ production in protonic and proton-nucleus collisions, for different existing fixed target experiments. In case of proton-nucleus collisions, our calculations show a  non-negligible dependence of the final state nuclear dissociation of J/$\psi$ mesons on the energy of the incident proton beam. The model has been applied to predict the J/$\psi$ production and suppression expected in proton-nucleus collisions at energies relevant to FAIR, the upcoming accelerator facility at Darmstadt, Germany. The amount of suppressions, for different mechanisms of  J/$\psi$ hadronization has been found to be distinguishably different which might help an experimental settlement of the much controversial issue of color neutralization.

\end{abstract}

\maketitle
\section{Introduction}

Collision of heavy nuclei at relativistic energies endows us with the unique opportunity to create and investigate hot and dense nuclear matter in the laboratory. Statistical calculations of quantum chromodynamics (QCD), the fundamental theory governing the dynamics of strongly interacting particles predict that at sufficiently high temperature and/or density nuclear matter undergoes a transition from the confined hadronic phase to a color de-confined partonic phase. Hence the primary objective of the relativistic heavy-ion physics program is the exploration of the phase diagram of the strongly interacting matter, in various regions of temperature and baryon density and to confirm the existence of this novel phase of quark matter, the quark-gluon plasma (QGP). Over the last three decades relentless efforts have thus been invested to devise unambiguous and experimentally viable signals that would clearly identify the existence of this phase transition and the signatures of this transient phase. The suppression of charmonia in high energy heavy-ion collisions, has long been predicted as a 'smoking gun' signature for the de-confinement transition \cite{MS,Satz,Vogt}. More than 25 years ago, Matsui and Satz argued in their seminal paper, that the production yield of J/$\psi$ mesons, a bound state made up of charm anti-charm pair, should be considerably suppressed by Debye screening of the colored partons if relativistic collisions would result in the formation of de-confined QCD medium \cite{MS}. However subsequent experimental investigations have revealed a considerable reduction of the charmonium production already present in proton-nucleus (p-A) collisions compared to the hadronic collisions (scaled with binary nucleon-nucleon (NN) collisions) \cite{NA3,E772,NA38,NA50-200,NA50-400,NA50-450,E866,HERAB,NA60} . A complete understanding of this 'normal nuclear suppression' is mandatory to establish a robust unambiguous baseline reference , with respect to which the 'anomalous suppression', pertaining to formation of a de-confined partonic medium can be clearly identified.

The upcoming compressed baryonic matter (CBM) experiment at the Facility for Anti-proton and Ion Research (FAIR) aims at the exploration of the QCD phase diagram in the region of high baryon densities and moderate temperatures, by colliding light and heavy-ions in the beam energy range $E_{Lab}$ = 10 - 40 AGeV \cite{CBM}. Model calculations performed so far \cite{density} , based on transport as well as hydrodynamical equations, indicate that for central (b=0) collisions highest net baryon densities $\rho_{B}$ = 6 - 12 $\rho_{0}$ (where $\rho_{0}$ is the density of normal nuclear matter) are expected to be produced at the center of the collision zone. The experiment will thus provide us the unique  possibility to create and investigate extremely dense nuclear matter in the laboratory through measurements of bulk as well as rare probes. Such measurements will enable us to have deeper insight on some of the fundamental but yet enigmatic issues of strong interaction thermodynamics like, hadronic properties in dense nuclear matter, de-confinement phase transition from hadronic to quark-gluon matter driven by high baryon densities and the nuclear equation of state (EOS) at high baryon densities. The charmonium produced at an early stage of the collision is believed to be a penetrating probe for characterization of the confining status of the dense baryonic medium. The experimental program thus includes a detailed proposal to study the production of the hidden charm hadrons, through their decay into lepton pairs \cite{Peter}, by high intensity beams. The di-leptons are not affected by the strong interactions that reign during hadronization and carry undistorted images of the earlier phases. At FAIR, production of charmed particles will be studied at energies close to their kinematic threshold (the threshold beam energy for J/$\psi$ production in p-p collisions being E$_{th}$ $\sim$ 11.3 GeV) and hitherto unexplored by the existing experiments, due to extremely low production cross sections.

In this paper we present a detailed study of the charmonium production in proton-proton (p-p) as well as proton-nucleus (p-A) collisions and its observed suppression in case of nuclear targets. To account for the nuclear reduction of the J/$\psi$ yield, some of the so called 'cold nuclear matter' (CNM) effects affecting the production in p-A collisions, have been incorporated in the model. Results from our calculations have been compared and contrasted with the data for J/$\psi$ production at fixed target energies. The outcomes of our investigation have been utilized to predict the J/$\psi$ production and its suppression expected to occur in the FAIR energy domain.

The paper is organized in the following way. The model for calculation of J/$\psi$ production in hadronic as well as nuclear collisions is briefly depicted in section II. Section III describes the analysis of  the existing J/$\psi$ production cross section data in p-p and p-A collisions, available from different fixed target experiments. In section IV we present our results for the estimation of J/$\psi$ yield expected in energetic collisions at FAIR. Finally in section V we summarize our results.

\section{THEORETICAL FRAMEWORK FOR CALCULATION OF J/$\psi$ PRODUCTION}

In hadronic collisions, one parton (quark or gluon) from each of the colliding hadrons interacts strongly to produce a $c\bar{c}$ pair. Subsequently the $c\bar{c}$ pair hadronizes to form a physical J/$\psi$ meson, provided the invariant mass of the pair is below the threshold for open charm production. Our present work has adopted the model originally proposed by Qiu, Vary and Zhang \cite{Qui} and subsequently used by others \cite{AKC} for interpretation of J/$\psi$ suppression in nucleon-nucleus and nucleus-nucleus collisions at SPS and RHIC. In this model, the J/$\psi$ production in high energy hadronic collisions is believed to a factorisable two stage process. The first stage is the production of a $c\bar{c}$ pair with relative momentum square $q^2$. Because of large charm quark mass, the process involves a short time scale and can be well accounted by perturbative QCD. The second stage involves the formation of the color neutral physical bound state J/$\psi$, from the initially compact $c\bar{c}$ pair. It takes a relatively longer time and thus non-perturbative in nature. Over the years, the main debate on the mechanism of J/$\psi$ production is focused on this second stage, which is not yet fully understood on a fundamental theoretical level. Three models are commonly found in literature for calculating the cross sections of J/$\psi$ production: the Color Evaporation Model (CEM) \cite{Evap}, the Color Octet Model (COM) \cite{Octet} and the Color Singlet Model (CSM) \cite{Singlet}. The production rate of the $c\bar{c}$ pairs with an invariant mass $Q^2$, can be factorized into i) a convolution of two parton distributions from the two incoming hadrons and, ii) $d\hat{\sigma}_{a+b\rightarrow c\bar{c}+X}/dQ^2$, which represents the perturbatively calculable short-distance hard parts for the parton $a$ and $b$ to produce the $c\bar{c}$ pairs with mass $Q^2$ \cite{CSS}. At the leading order, the partonic contributions come from two subprocesses: quark annihilation ($q\bar{q} \rightarrow c\bar{c}$) and gluon fusion ($gg\rightarrow c\bar{c}$).  With the K-factor accounting for effective higher order contributions, the single differential J/$\psi$ production cross section in collisions of hadrons  $h_1$ and $h_2$, at the center of mass energy $\sqrt{s}$ can be expressed as,

\begin{equation}
\label{diff}
\frac{d\sigma_{h_1h_2}^{J/\psi}}{dx_F} = K_{J/\psi}\int dQ^{2}\left(\frac{d\sigma_{h_1h_2}^{c\bar{c}}}{dQ^2dx_F}\right)\times F_{c\bar{c}
\rightarrow J/\psi}(q^2), 
\end{equation}

\noindent  where $Q^2 = q^2 +4 m_C^2$ with $m_C$ being the mass of the charm quark and $x_F$ is the Feynman scaling variable. It can be related to $x_a$ and $x_b$ by $x_F = x_a - x_b$, with $x_a$ and $x_b$ being the momentum fractions carried by the incoming partons originating from the beam and the target respectively. In  Eq.~(\ref{diff}) $F_{c\bar{c} \rightarrow  J/\psi}(q^2)$  is  the transition probability that a $c\bar{c}$ pair with relative momentum square $q^2$ evolve into a physical  $J/\psi$  meson, in hadronic collisions.  Three  different parametric forms have been formulated for the transition probability corresponding to the three above mentioned physical mechanisms of color neutralization:

\begin{mathletters}

\begin{eqnarray}
F^{\rm (C)}_{c\bar{c}\rightarrow{\rm J/}\psi}(q^2)
&=& N_{{\rm J/}\psi}\, \theta(q^2)\, \theta(4m_D^2-4m_C^2-q^2) \ ,
\label{const}
\\
F^{\rm (G)}_{c\bar{c}\rightarrow{\rm J/}\psi}(q^2)
&=& N_{{\rm J/}\psi}\, \theta(q^2)\,
\exp\left[-q^2/(2\alpha^2_F)\right] \ ,
\label{gauss}
\\
F^{\rm (P)}_{c\bar{c}\rightarrow{\rm J/}\psi}(q^2)
&=& N_{{\rm J/}\psi}\, \theta(q^2)\, \theta(4m_D^2-4m_C^2-q^2)
\nonumber \\
&\times & \left(1-q^2/(4m_D^2-4m_C^2)\right)^{\alpha_F}\ ,
\label{power}
\end{eqnarray}
\end{mathletters}
where $m_D$ is the mass scale for the open charm production threshold ($D\bar{D}$ threshold). In $F_{c\bar{c} \rightarrow  J/\psi}(q^2)$, $N_{{\rm J/}\psi}$ and $\alpha_F$ are the model parameters that can be fixed by comparing the model results with the existing total production cross section data from hadron-hadron collisions.

As illustrated in \cite{Qui}, the different functional forms of the transition probability represent different pictures of color neutralization. The $F^{\rm (C)}(q^2)$ implies that a constant fraction of all the produced $c\bar{c}$ pairs with invariant mass below the open charm production threshold evolves into the physical J/$\psi$ mesons, and thus bears the central theme of the Color Evaporation Model \cite{Evap}. $F^{\rm (G)}(q^2)$ carries the essential ingredients of the Color-Singlet Model \cite{Singlet}. It is assumed that the $c\bar{c}$ pair is produced in a color-singlet state. The transition amplitude $\langle c\bar{c} | {\rm J/}\psi\rangle$ does not involve any radiation and interaction with the medium, and thus proportional to square of the J/$\psi$ wave function parameterized as Gaussian. Finally, $F^{\rm (P)}(q^2)$ in Eq.~(\ref{power}), mimics the the essential features of Color-Octet Model \cite{Octet}. In this model the $c\bar{c}$ pairs are assumed to be produced in color-octet stage. Subsequent formation of color-singlet physical resonances occurs through the gradual expansion of the initially compact $c\bar{c}$ pairs. The expansion is associated with the radiation of soft gluons necessary for color neutralization. The $q^2$-dependence of the transition probability is assumed to be associated with that radiation, and a power-law (P) distribution, is believed to represent the transition probability.

Instead of $x_F$, one can also express the single differential cross section, in terms of the center of mass rapidity $y_{cms}$, of the $c\bar{c}$ pair as,

\begin{equation}
\label{diffrap}
\frac{d\sigma_{h_1h_2}^{J/\psi}}{dy_{cms}} = (x_a+x_b) \frac{d\sigma_{h_1h_2}^{J/\psi}}{dx_{F}}
\end{equation}

The inclusive cross sections as reported by different experiments can then be estimated by integrating over corresponding kinematic range.

\begin{figure} \vspace{-0.1truein}
\includegraphics[width=8.6cm]{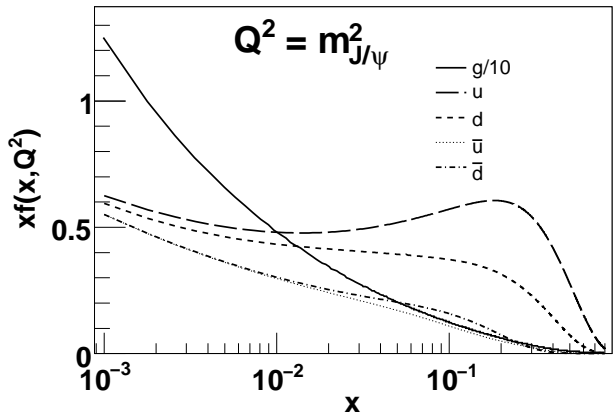}
\caption{\footnotesize The MSTW 2008 leading order (LO) parton distribution functions in a free proton for gluon (continuous line), up quark (long dashed), down quark (short dashed), anti-up quark (dotted line) and anti-down qurak (dashed-dot line). The parton densities are evaluated at a scale $Q=m_{J/\psi}$.}
\label{fig1}
\end{figure}

The double differential cross section for producing a $c\bar{c}$ pair can be decomposed into the individual contributions from different channels and at leading order is given by
\begin{equation}
\label{doublediff}
{d\sigma_{h_1h_2}^{c\bar c}\over dQ^2dx_F}={d\sigma^{q\bar
q}_{h_1h_2}\over dQ^2dx_F}
+{d\sigma^{c\bar c_{int}}_{h_1h_2}\over dQ^2dx_F}
+{d\sigma^{gg}_{h_1h_2}\over dQ^2dx_F},
\end{equation}

where $\sigma^{q\bar q}$ is the contribution from annihilation of light quarks, $q=u,d,s$, $\sigma^{c\bar{c}_{int}}$ is the contribution from intrinsic charm quarks, and $\sigma^{gg}$ is the contribution from fusion of two gluons. Due to very small charmed sea distribution appearing quadratically in the cross section calculation, the contribution of intrinsic component $\sigma^{c\bar{c}_{int}}$, is negligible compared to the other terms \cite{BQV}. Hence the significant contributions from the other two leading terms are

\begin{eqnarray}
\label{quark}
 &{d\sigma^{q\bar q}_{h_1h_2}\over dQ^2dx_F}=\sum_{f=u,d,s}
{\hat{\sigma}^{q\bar q}(Q^2)\over Q^2}{x_ax_b\over
x_a+x_b}\nonumber\\
& \times \left[{{{q_h}_1}^f}(x_a,Q^2)\bar {{{q_h}_2}^f}(x_b,Q^2)+\bar {{{q_h}_1}^f} (x_a,Q^2) {{{q_h}_2}^f}(x_b,Q^2)\right]
\end{eqnarray}

for light quark annihilation, and

\begin{equation}
\label{gluon}
{d\sigma^{gg}_{h_1 h_2}\over dQ^2dx_F}=
{\hat{\sigma}^{gg}(Q^2)\over Q^2}{x_a x_b\over x_a + x_b}
{{g_h}_1}(x_a,Q^2){{g_h}_2}(x_b,Q^2),
\end{equation}

for gluon fusion.  Here $q(x,Q^2)$, $\bar q(x,Q^2)$ and $g(x,Q^2)$ signify the quark, anti-quark and gluon distributions, respectively, in a hadron, evaluated at a scale $Q^2$. In the present work, we have used MSTW 2008 \cite {MSTW} LO parton distribution functions for free protons. Fig.~\ref{fig1}  shows the momentum distributions for  up and down quarks, anti-quarks and gluons following MSTW 2008 LO central set, evaluated at the scale $Q=m_{J/\psi}$. Because of the two-parton final-state at the leading order, the incoming parton momentum fractions are fixed by the kinematics, and at a given $\sqrt{s}$, can be expressed in terms of $x_F$ and $Q^2$, by $x_a =(\sqrt{x_F^2 + 4Q^2/s} + x_F)/2$ and $x_b =(\sqrt{x_F^2 + 4Q^2/s} - x_F)/2$, respectively. The partonic cross sections appearing in Eqs.~(\ref{quark}) and (\ref{gluon}), in the leading log approximation, are given by \cite{BQV,cross}
\begin{equation}
\label{anni}
\hat{\sigma}^{q\bar q}(Q^2)={2\over 9}
{4\pi\alpha^2_s\over 3Q^2}(1+{1\over
2}\gamma)\sqrt{1-\gamma},
\end{equation}
and
\begin{eqnarray}
\hat{\sigma}^{gg}(Q^2)&=&{\pi\alpha^2_s\over 3Q^2}
\big{[}(1+\gamma+{1\over 16}\gamma^2)
\log({1+\sqrt{1-\gamma}\over
1-\sqrt{1-\gamma}})\nonumber\\
& &\,\,\,-({7\over 4}+{31\over
16}\gamma)\sqrt{1-\gamma}\,\big{]},
\label{fusion}
\end{eqnarray}
where $\alpha_s$ is the QCD running coupling constant, and $\gamma=4m_c^2/Q^2$.

Let us now turn to a description of J/$\psi$ production in proton-nucleus collisions. In this case, the presence of normal nuclear matter can affect the charmonium production and thus these collisions provide a tool to probe the effect of confined matter. Nuclear effects can come into play during the entire evolution period of J/$\psi$ production. Several different phenomena collectively known as cold nuclear matter (CNM) effects have been studied in literature at considerable detail. Two most important CNM effects that have been identified are: i) the initial state nuclear modification of the parton distribution function affecting the perturbative $c\bar{c}$ pair production, ii) the final state dissociation of the nascent $c\bar{c}$ pair in the pre-resonance  or resonance stage, due to its interactions with nucleons during its passage through the target nucleus.

The PDF in a nucleus with atomic number $Z$ and mass number $A$ is written as the sum of the proton ($f_{i}^{p/A}$) and the neutron ($f_{i}^{n/A}$) parton densities in a nucleus:
\begin{equation}
  \label{isospin}
  f_{i}^{A} = Z \ f_{i}^{p/A} + (A-Z) \ f_{i}^{n/A},
\end{equation}

where $f_{i}^{n/A}$ is obtained from $f_{i}^{p/A}$ by isospin conjugation: $u^{n/A}=d^{p/A}$, $d^{n/A}=u^{p/A}$, $s^{n/A}=s^{p/A}$. Deep inelastic scattering (DIS) and Drell-Yan measurements performed with nuclear targets have shown that the distributions of partons in nuclei are significantly modified relative to those in free protons. These nuclear modifications depend on the fraction of the total hadron momentum carried by the parton, $x$, on the momentum scale, $Q^2$, and on the mass number of the nucleus, $A$.  While the mechanisms governing these modifications are not yet well understood, several groups have produced parameterizations, $R_i(A,x,Q^2)$, that convert the free-proton distributions for each parton $i$, $f_i^{p}(x,Q^2)$, into nuclear ones, $f_i^A(x,Q^2)$, assuming factorization:
\begin{equation}
f_i^A(x,Q^2) = R_i(A,x,Q^2) \times f_i^{\rm p}(x,Q^2) \quad .
\label{npdfs}
\end{equation}

\begin{figure} \vspace{-0.1truein}
\includegraphics[width=8.6cm]{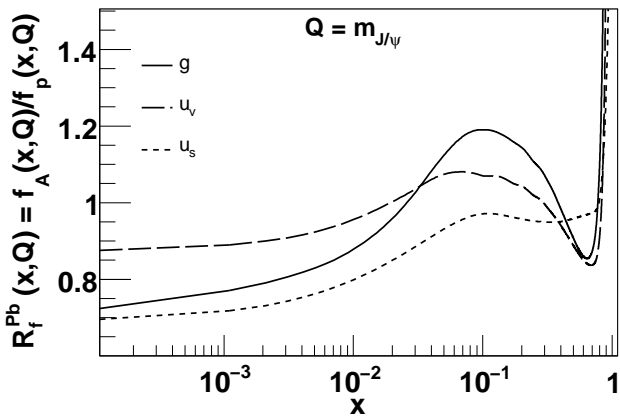}
\caption{\footnotesize Nuclear modification of valence up quark (long dashed line), sea up quark (short dashed line) and gluon (continuous line) distribution inside a Pb nucleus following EPS09 interface evaluated  at the scale $Q = m_{J/\psi}$. EPS09 gives $R_i(Pb,x,Q^2)$, the scale dependent ratio of the distribution of the parton flavor i for a proton in a nucleus A, to the corresponding parton distribution in the free proton.}
\label{fig2}
\end{figure}

 For our calculations, we have employed recently proposed EPS09 \cite{EPS09} interface, which is available for all mass numbers. Fig.~\ref{fig2} shows the ratio $R_i(A,x,Q^2)$ following EPS09 parameterization, in a Pb nucleus for valence up quarks, sea up quarks and gluons, calculated at the scale, $Q = m_{J/\psi}$. It is important to note that, the quark and anti-quark distribution functions are directly probed by the nuclear DIS and Drell-Yan data, their nuclear effects are relatively well constrained and different parameterizations give almost similar results.  The connection between the measurements and the nuclear gluon densities is much more indirect. Since gluon fusion plays the dominant role for $c\bar{c}$ production, a complete understanding of charmonium production in proton-nucleus collisions is presently lacking due to uncertainties in nuclear gluon distributions, as illustrated in \cite{Lourenco} . It may be mentioned here that in the original model by Qui, Vary and Zhang the nuclear effect to the parton distribution functions are ignored and free proton pdfs have been used to estimate the $c\bar{c}$ pair production in nucleon-nucleus and nucleus-nucleus collisions.

For simulating the final state interactions we closely follow the method advocated by Qui, Vary and Zhang \cite{Qui,BQV}, who have treated the conventional nuclear dissociation in somewhat unconventional way. In nucleon-nucleus collisions, the produced $c\bar{c}$ pairs are likely to interact with the nuclear medium before they exit. Observed anomalous nuclear enhancement of the momentum imbalance in di-jet production \cite{QVZ} indicates that a colored parton (quark or gluon) experiences multiple soft scatterings when it exits the target nucleus \cite{LQS}. Hence, the colored $c\bar{c}$ pair produced through partonic hard scattering should also experience multiple soft scatterings when it passes through nuclear matter. These multiple soft scatterings will increase the relative transverse momentum between the $c$ and $\bar c$, and consequently, increase the invariant mass of the $c\bar{c}$ pair. Some  pairs can thus gain enough momentum to be pushed over the threshold and  become two open charm mesons, and the J/$\psi$ production cross section will in turn be reduced in comparison with nucleon targets. Larger be the size of the nucleus, more be the number of soft scatterings undergone by the pair and the reduction in the cross section for J/$\psi$ production will be even greater. If the formation length for the J/$\psi$ meson, which depends on the momenta of the $c\bar{c}$ pairs produced in the hard collision, is longer than the size of the nuclear medium, it is reasonable to assume that the transition probability $F_{c\bar{c}\rightarrow {\rm J/}\psi}(q^2)$, defined in Eq.~(\ref{diff}), can be factorized from the multiple scattering. Then, as far as the total cross section is concerned, the net effect of the multiple scattering of the $c\bar{c}$ pairs can be represented by a shift of $q^2$ in the transition probability, 
\begin{equation}
q^2  \longrightarrow  \bar{q}^2 = q^2 + \varepsilon^2\, L(A) \ .
\label{q2shift}
\end{equation}

In Eq.~(\ref{q2shift}), $L(A)$ is the effective length of nuclear medium traversed by the $c\bar{c}$ pair, from its point of production till it exits and depends on the details of the nuclear density distributions \cite{LABref}. In Eq.~(\ref{q2shift}), $\varepsilon^2$ represents the square of the relative momentum received by the $c\bar{c}$ pairs per unit length of the nuclear medium.

\section{ANALYSIS OF J/$\psi$ PRODUCTION CROSS SECTIONS AT FIXED TARGET ENERGIES}

\begin{figure*} \vspace{-0.1truein}
\includegraphics[width=15.6cm]{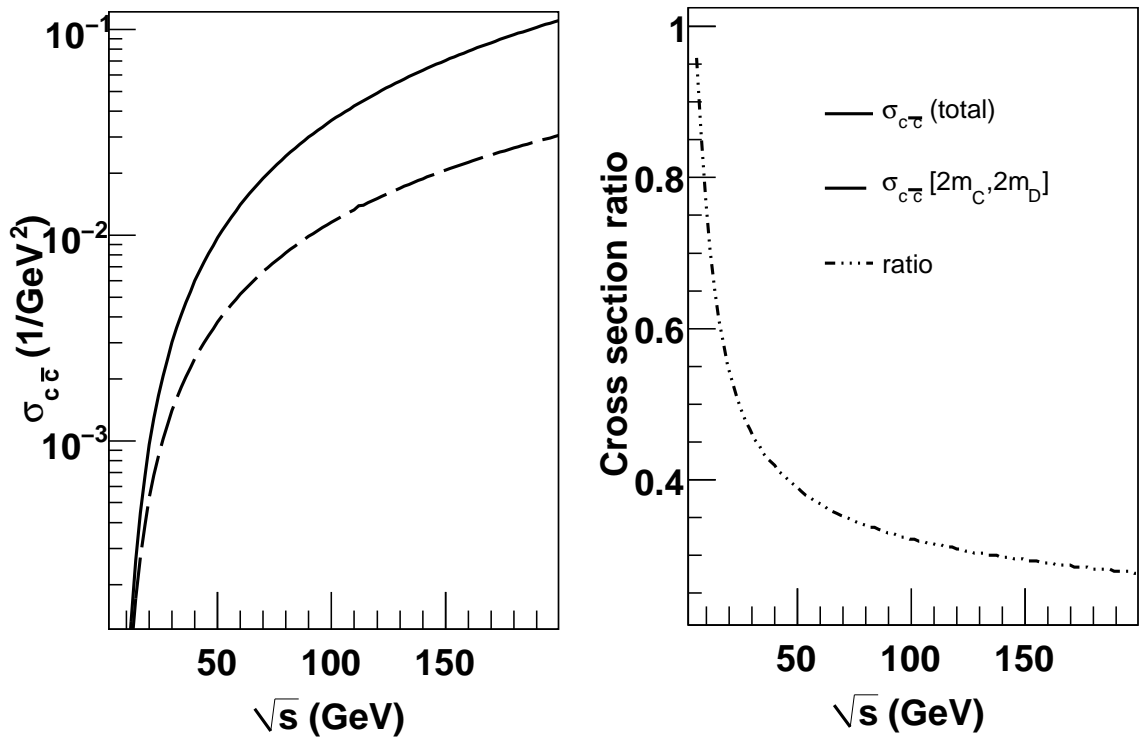}
\caption{\footnotesize Variation of the sub-threshold (dashed line) and total (continuous line) $c\bar{c}$ production (left panel) and their ratio (sub-threshold fraction) (right panel) as a function of the center of mass energy of the colliding hadrons.}
\label{fig3}
\end{figure*}

In this section, we will describe the results of our analysis of J/$\psi$ production cross sections in hadronic as well as hadron-nucleus collisions. Models describing quarkonium production in hadronic collisions following color-octet or color evaporation approach generally assume that the perturbatively produced $c\bar{c}$ pairs can subsequently hadronize into observed hidden charm mesons (with a certain probability) provided the invariant mass of the pair lies in the window $2m_C \le Q \le 2m_D$. Thus it might be interesting to study the energy evolution of these sub-threshold $c\bar{c}$ pairs. In Fig.~\ref{fig3} we have thus plotted  sub-threshold $\sigma_{c\bar{c}} [2m_{C}, 2m_{D}]$, and total $c\bar{c}$ production cross section and their ratio as a function of the center of mass energy $\sqrt{s}$ of the colliding hadrons. At lower energies the two cross sections are almost indifferent indicating that all the $c\bar{c}$ pairs are predominantly produced within the hidden charm interval. However the sub-threshold cross section is found to rise less steeply with energy compared to the total cross section. The ratio of these two cross sections, the sub-threshold fraction is first seen to decrease with energy then gradually saturates for large $\sqrt{s}$. Thus higher be the energy of the collision, $c\bar{c}$ pairs are more likely to be produced with an invariant mass above the open charm production threshold.

\begin{figure*} \vspace{-0.1truein}
\includegraphics[width=8.5cm]{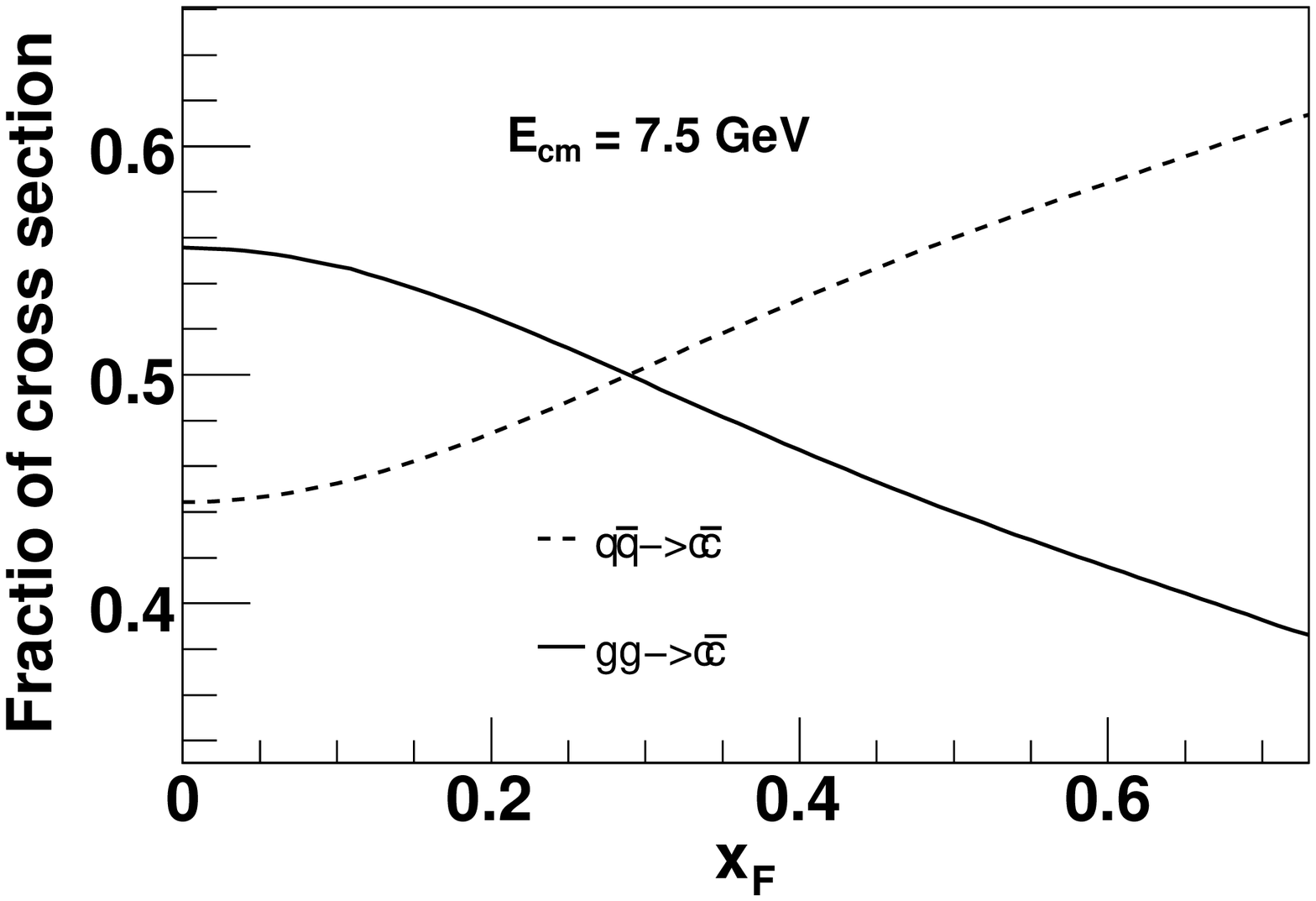}
\includegraphics[width=8.5cm]{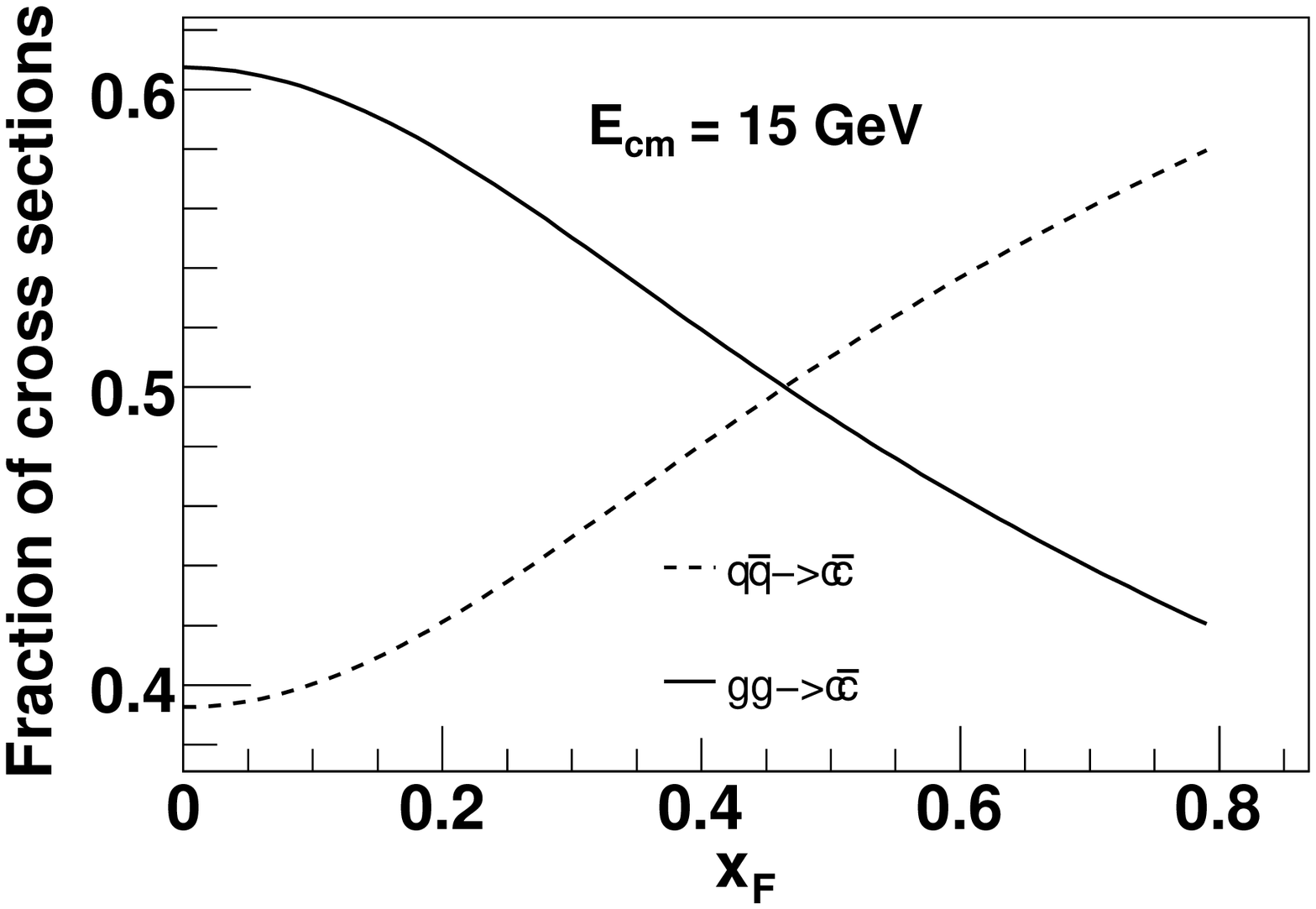}
\includegraphics[width=8.5cm]{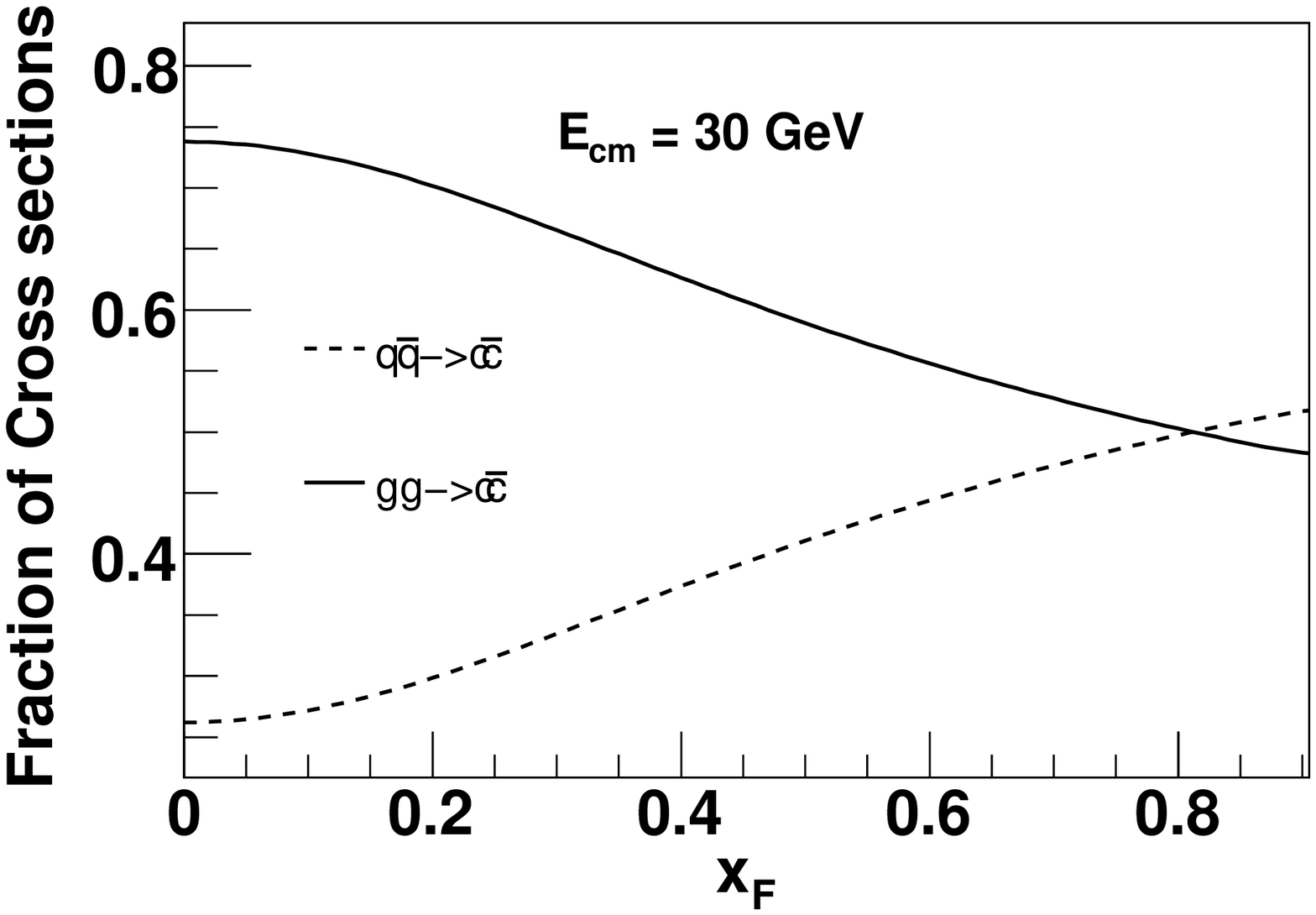}
\includegraphics[width=8.5cm]{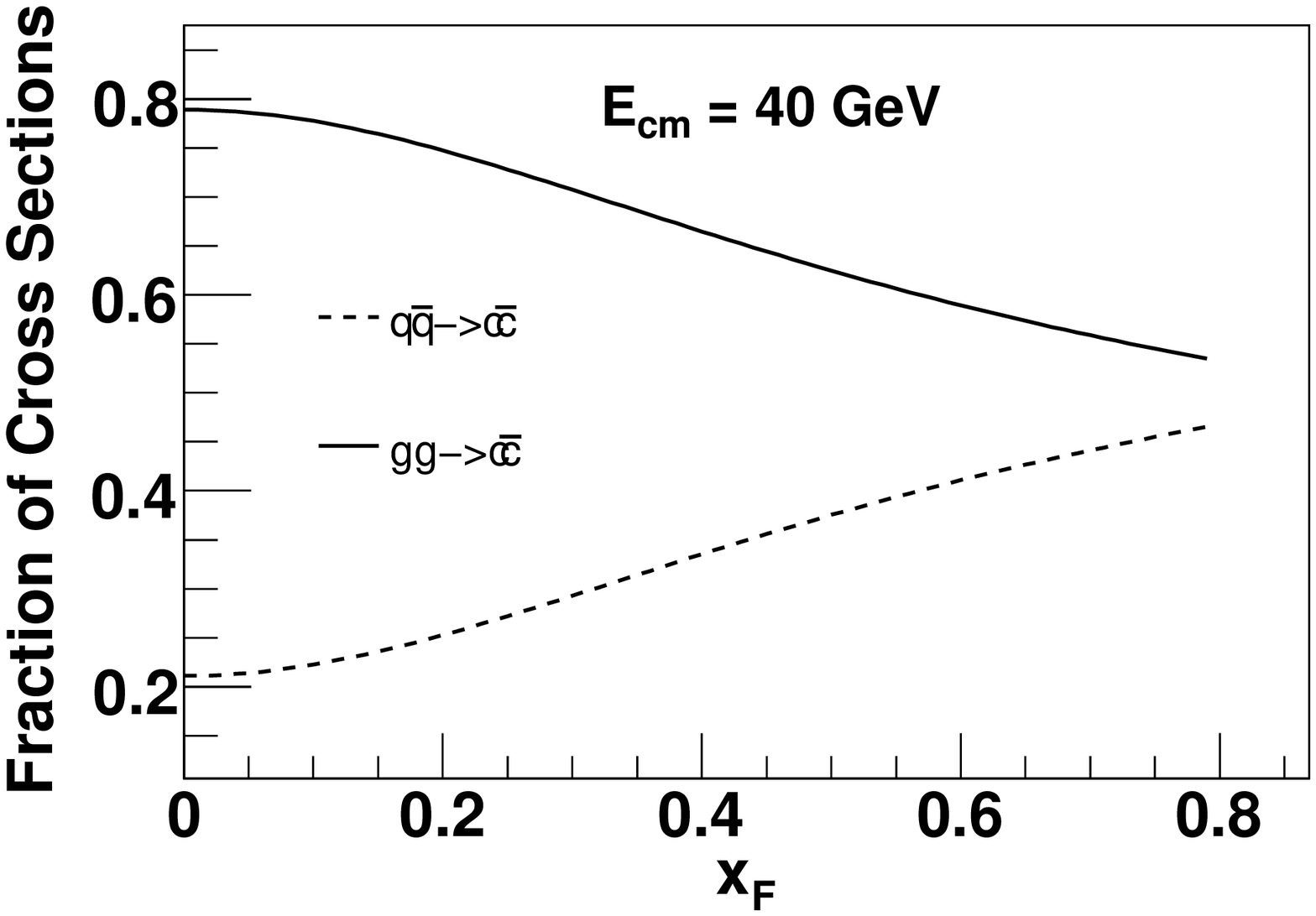}

\caption{\footnotesize Ratio of light quark annihilation and gluon fusion cross sections for $c\bar{c}$ production in p-p collisions to their sum as a function of the Feynman scaling variable $x_F$, at different values of the center of mass energies of the collision system. MSTW 2008 LO central pdf set is used as an input.The solid line is for quark contribution , and the dashed line is for gluonic contribution as indicated in the figures.}
\label{fig4}
\end{figure*}

As pointed out earlier, at leading order, the significant contribution to $c\bar{c}$ production comes from  the light quark annihilation and gluon-gluon fusions. Because of the larger color factors, the inclusive cross section, at a given energy, due to gluon fusion is always higher than that due to quark annihilation. However this is not the case if one looks at the differential cross sections of these two processes. To estimate their relative size, we plot in Fig.~\ref{fig4}, the fraction of the total differential cross section due to each sub-process in p-p collision for different values of  $\sqrt{s}$.  At lower $x_F$, gluon fusion dominates the $c\bar{c}$ production whereas at higher $x_F$, quark annihilation also plays a crucial role. The critical value $x_F^{Crit}$, above which $q\bar{q}$ annihilation cross section rides over gg fusion is not fixed and depends on the underlying collision energy. Higher be the $\sqrt{s}$, larger is the contribution due to gluons. Consequently at RHIC ($\sqrt{s_{NN}} = 200$ GeV) or at LHC ($\sqrt{s_{NN}} = 5.5$ TeV), J/$\psi$ production up to leading order can be reliably described by gg fusion alone. However, at the other end of the energy scale, in the FAIR energy domain, one needs to consider the light quark annihilation channel as well. This observation can be easily understood by looking at the behavior of  parton distribution functions shown in  Fig.~\ref{fig1}. Larger $x_F$ requires a beam parton at large $x_a$, which in turn means small $\sqrt{s}$. In this region the gluon distribution of a proton is very small compared to the valence quark distribution. Consequently, the J/$\psi$ production cross section at large $x_F$ is very sensitive to the sea quark distributions of the target.

\begin{table}
\caption{Values of fitting parameters $f_{J/\psi}$ and $\alpha_F$ used in the calculation of J/$\psi$ production cross sections}
\label{table1}
\begin{tabular}{|c|cc|cc|cc|}
\hline
 & $F^{\rm (C)}$ & & $F^{\rm (G)}$ & & $F^{\rm (P)}$ &
\\ \hline
$f_{{\rm J/}\psi}$&  0.261        & &    0.517      & & 0.510         &
\\ \hline
 $\alpha_F$       &   0           & &   1.2 GeV     & & 1.0           &
\\ \hline
\end{tabular}
\end{table}

\begin{figure} \vspace{-0.1truein}
\includegraphics[width=8.6cm]{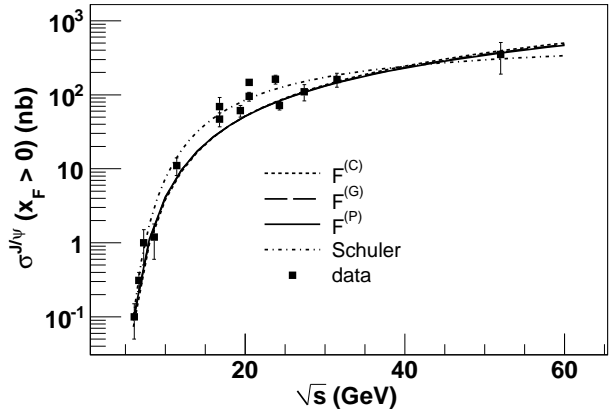}
\caption{\footnotesize Total J/$\psi$ production cross section in the forward region ($x_F > 0$) as a function of center of mass energy. In addition to p-p collisions, the plot also contains the available data points for normalized proton-nucleus collisions with light targets from different fixed target energies. The model parameters for each of the parametric forms of the transition probability $F(q^2)$ is however fixed from the p-p data. The cross section as obtained from Schuler parameterization is also included.}
\label{fig5}
\end{figure}

For a complete description of charmonium production, let us now fix the the parameters of the model, described in the previous section. For each of the parametric forms of the transition probability, combining the K-factor with the overall normalization constant  $N_{J/\psi}$, we are left with two free parameters $\alpha_F$ and $f_{J/\psi} = K_{J/\psi}N_{J/\psi}$. The optimum values of these two parameters, given in Table \ref{table1}, are extracted by fitting the model results with the inclusive J/$\psi$ production cross section data in the forward region ($x_F > 0$), from pp collisions at fixed target energies. All the three parameterizations of $F_{c\bar{c} \rightarrow  J/\psi}(q^2)$ are independently found to fit the available data reasonably well. Results are shown in Fig.~\ref{fig5}. In addition to p-p reactions, data from normalized p-N collisions are also plotted. These nuclear target data points represent the forward cross section per nucleon from light targets (Li, Be, C). Their close agreement with the theoretical curves for p-p collisions indicates that for light-ions J/$\psi$ production is least affected by the nuclear effects due to the presence of few target nucleons. A parameterization, called Schuler parameterization \cite{Vogt} for total forward cross section not including the branching ratio to lepton pairs is also available from the E672 collaboration, for $\sqrt{s} \le 31 GeV$, as 

\begin{equation}
\label{Schuler}
{\sigma(x_F>0)}={\sigma}_{0}(1 - {m \over {\sqrt{s}}})^{n}
\end{equation}

where n = 12.0 $\pm 0.9$ and $\sigma_0$ = 638 $\pm 104$ nb for a proton beam. In Fig.~\ref{fig5}, we have also compared our results with this parameterization. 

\begin{figure} \vspace{-0.1truein}
\includegraphics[width=8.6cm]{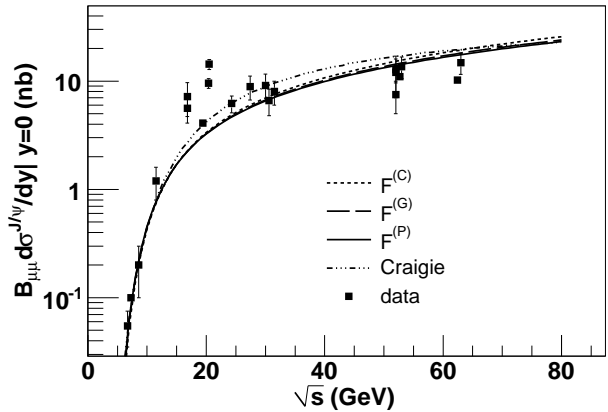}
\caption{\footnotesize Energy evolution of the mid-rapidity J/$\psi$ production cross section in di-muon channel for proton-nucleon collisions. The Craige parameterization of the differential cross section is also shown. Model calculations are performed following Eq.~\ref{diffrap}.}
\label{fig6}
\end{figure}

Once the parameters are fixed, the model can now also be used to calculate the energy dependence of the cross section at mid-rapidity. This energy dependence has also been parameterized long ago by Craigie \cite{Vogt} as:

\begin{equation}
\label{Craige}
B{d\sigma \over dy}=
    Cexp({-14.7m \over {\sqrt{s}}}) 
\end{equation} 

where m is the mass of the particular resonance state, B is the branching ratio for di-muon channel and C(J/$\psi$)=40.0 nb. Fig.~\ref{fig6} thus shows the comparison of the data for the mid-rapidity J/$\psi$ production in p-N collisions at fixed-target energies with the model calculations along with the parametric curve.

Thus we have seen that with all three parametric forms of the transition probabilities, representing three possible physical scenarios of color neutralization, the model gives a reasonable description of the data for J/$\psi$ hadroproduction at the available fixed target energies, an observation in line with the previous studies \cite{Qui}. Here a few words about the selected data sets on hadroproduction of charmonia may be in order. Quarkonium production cross sections are usually reported either as integrated over forward region of phase space, ($x_F > 0$) in the center of mass or in the central rapidity region ($y_{cms}=0$). The measurements have been performed over a time period spanning more than 30 years. Over such a long period, several experimental techniques have been used and different input information were available during the time of measurements. Hence comparing results of different experiments on an equal footing requires an update of the published numbers  on several aspects. For example, charmonium branching ratio has changed with time and the treatments of nuclear effects are not homogeneous. The data here are taken from \cite{Vogt}, where all cross section data are adjusted with the same di-muon branching ratio ($B_{\mu\mu} = 0.0597 \pm 0.0025$). Also nuclear target data are corrected to pp by assuming $\sigma_{pA} = \sigma_{pp}A^{0.9}$.

\begin{figure*} \vspace{-0.1truein}
\includegraphics[width=8.5cm]{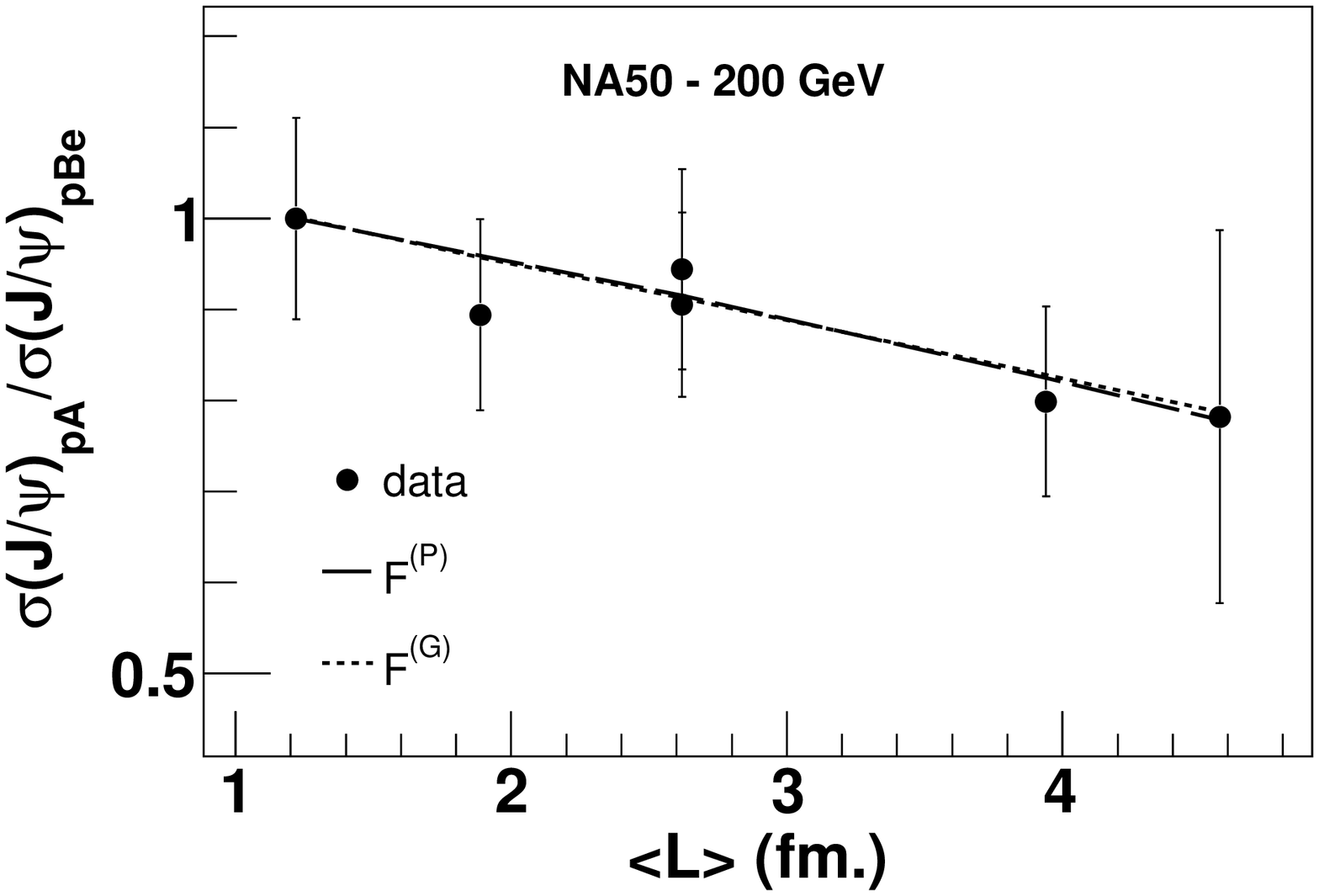}
\includegraphics[width=8.5cm]{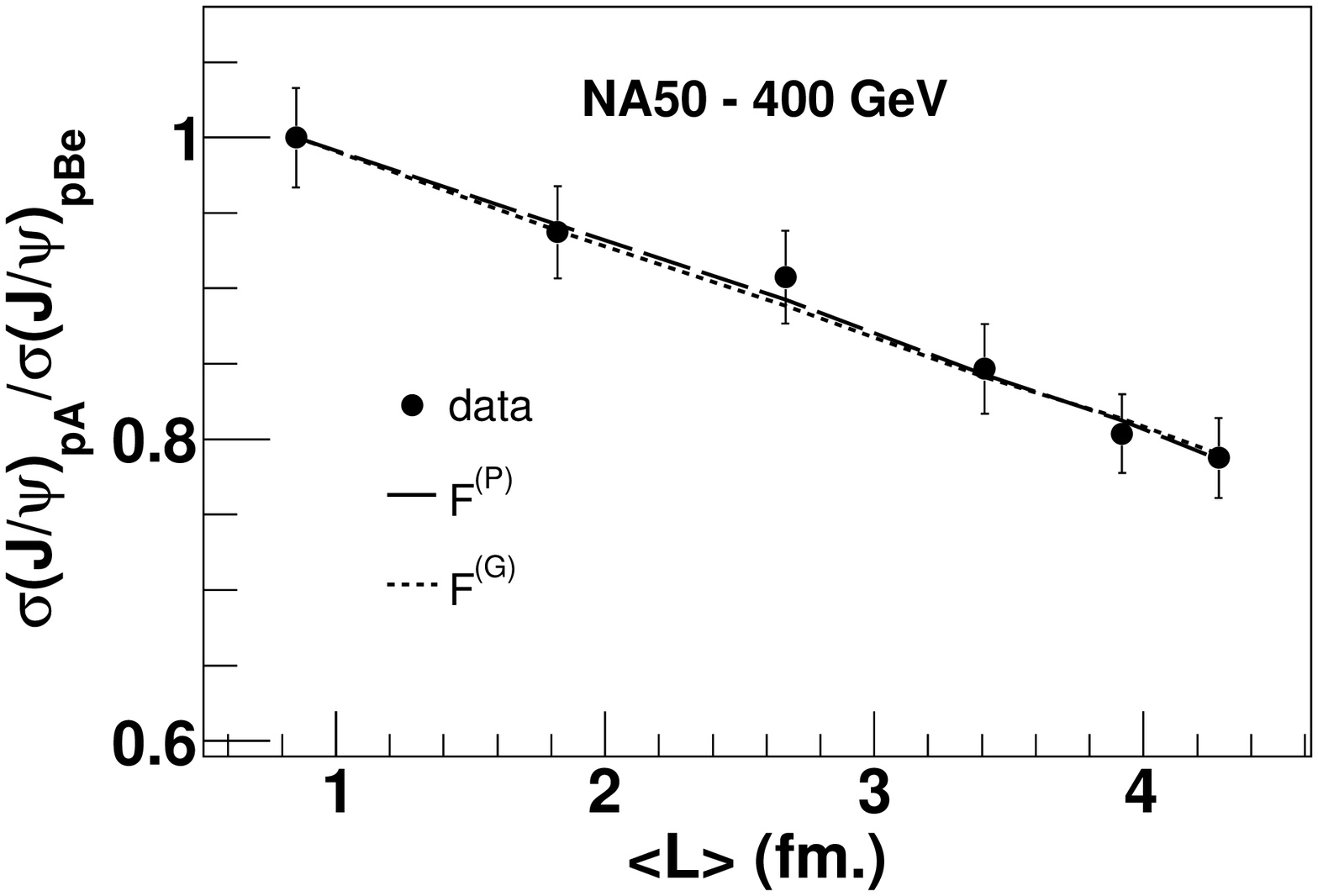}
\includegraphics[width=8.5cm]{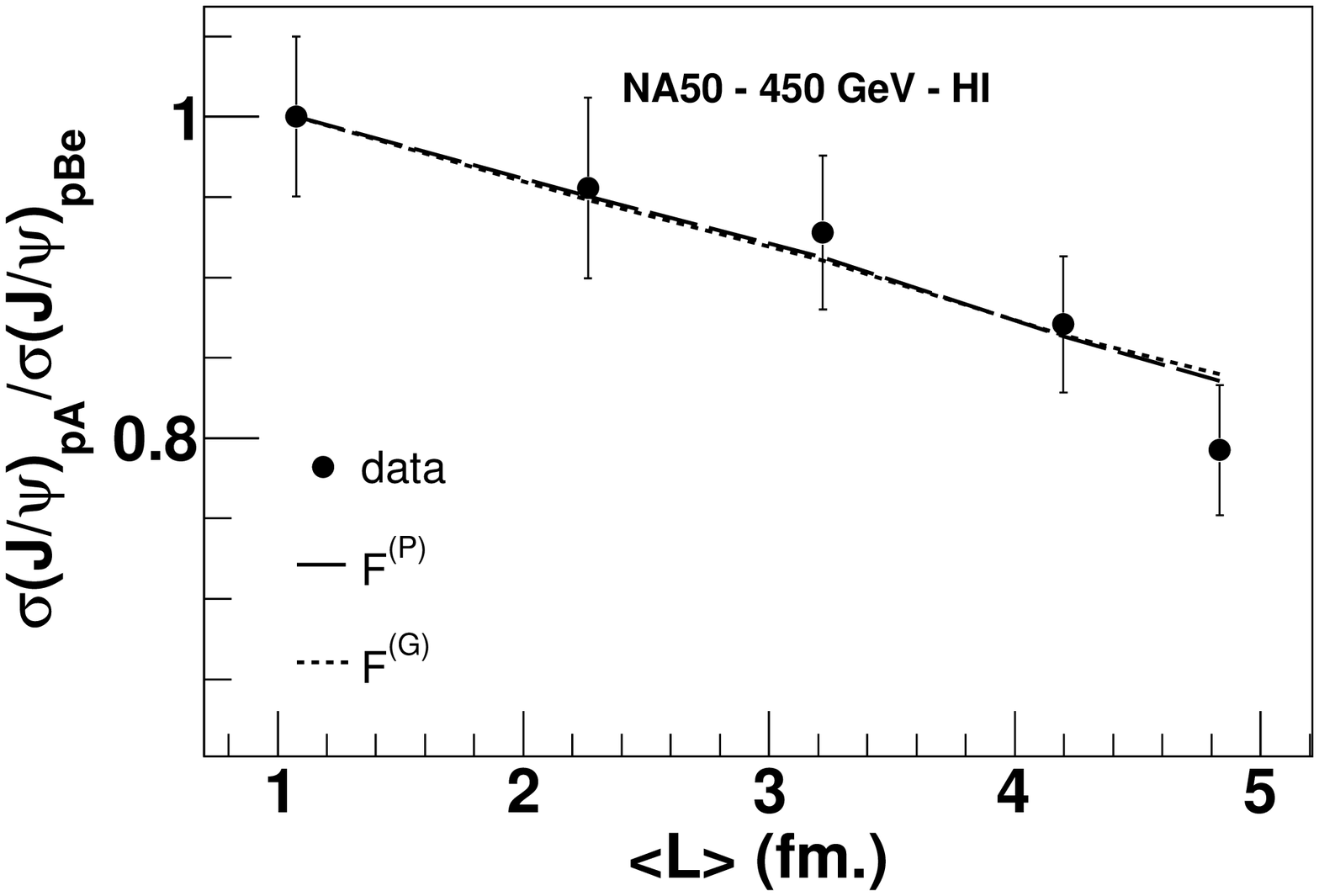}
\includegraphics[width=8.5cm]{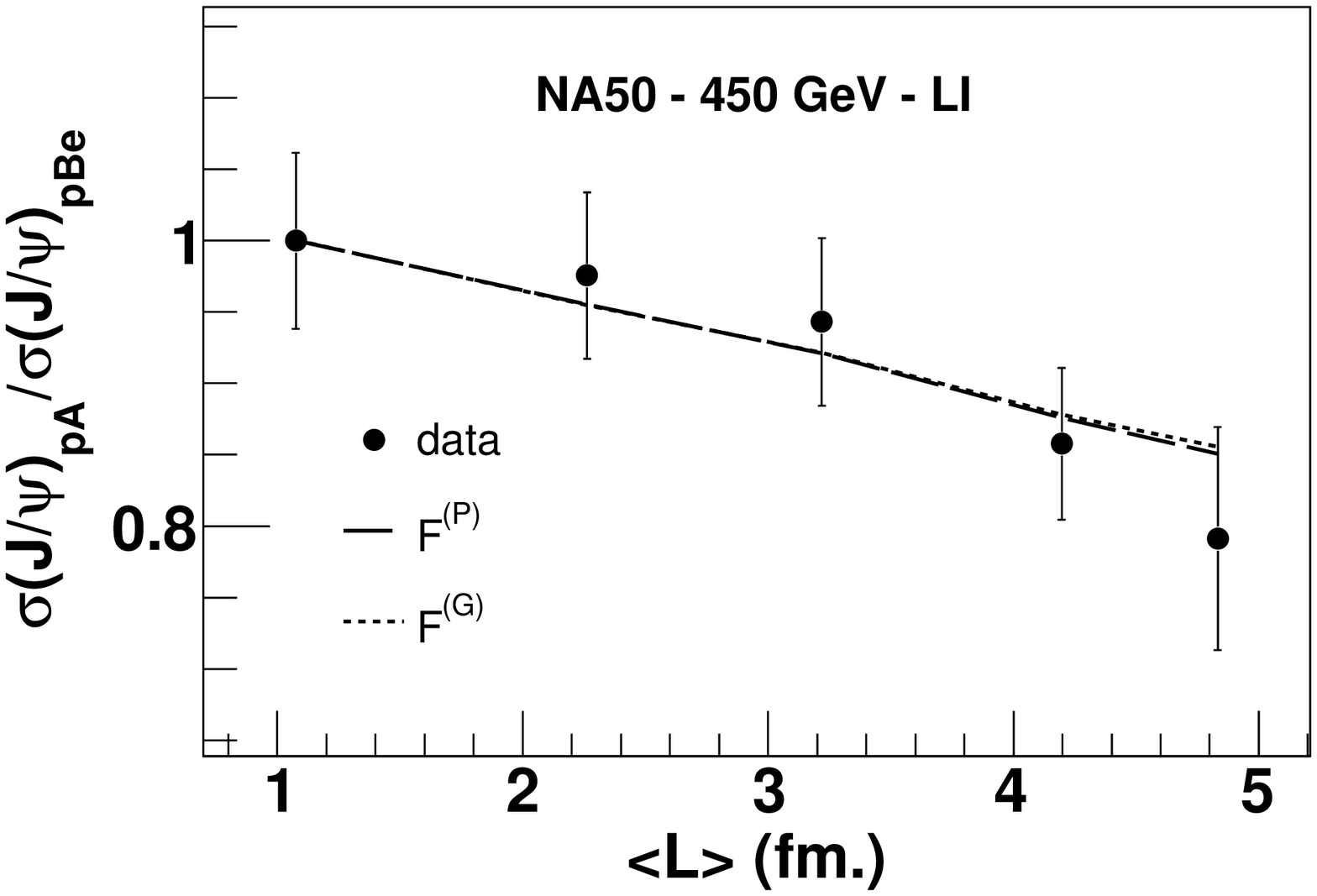}
\includegraphics[width=8.5cm]{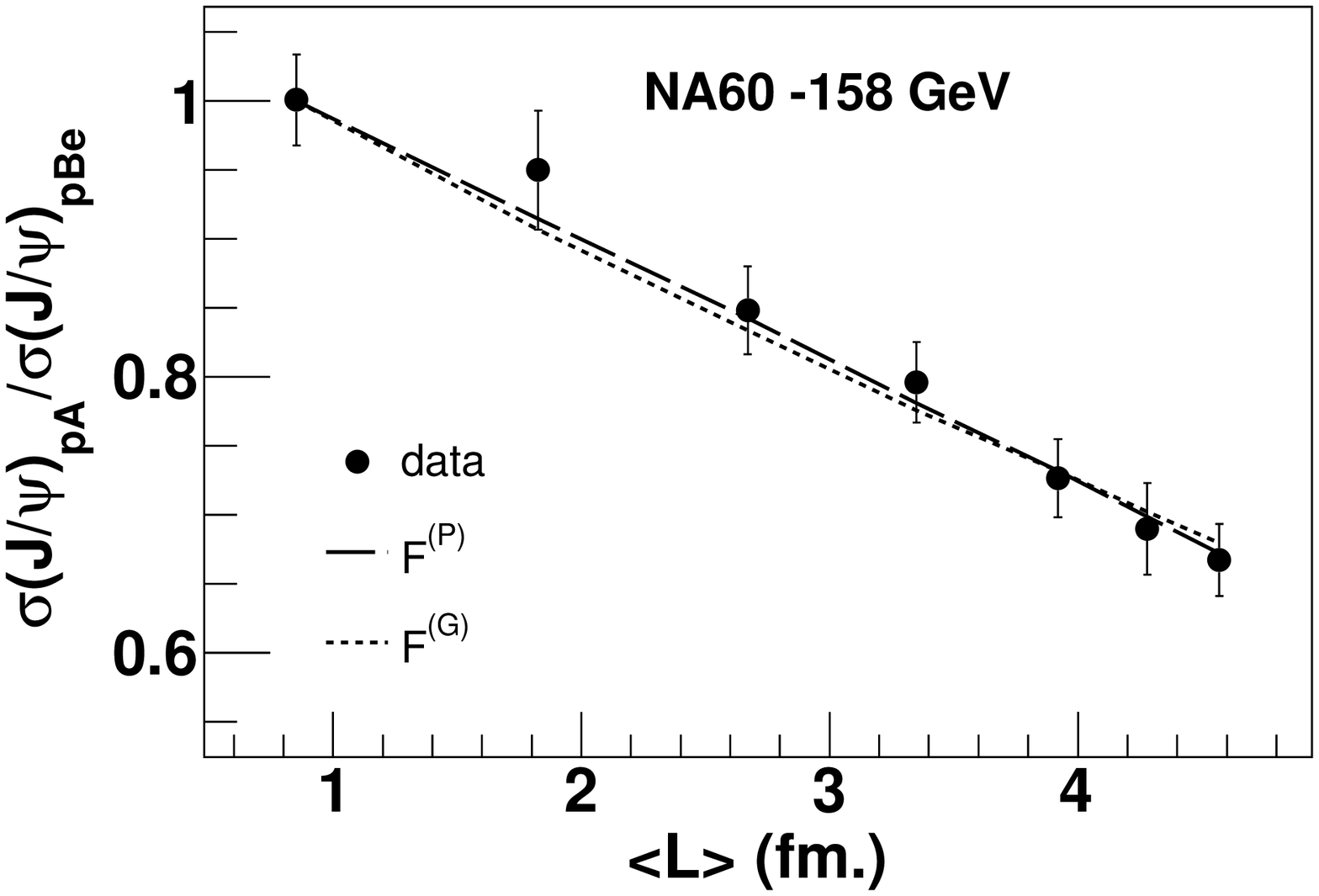}
\includegraphics[width=8.5cm]{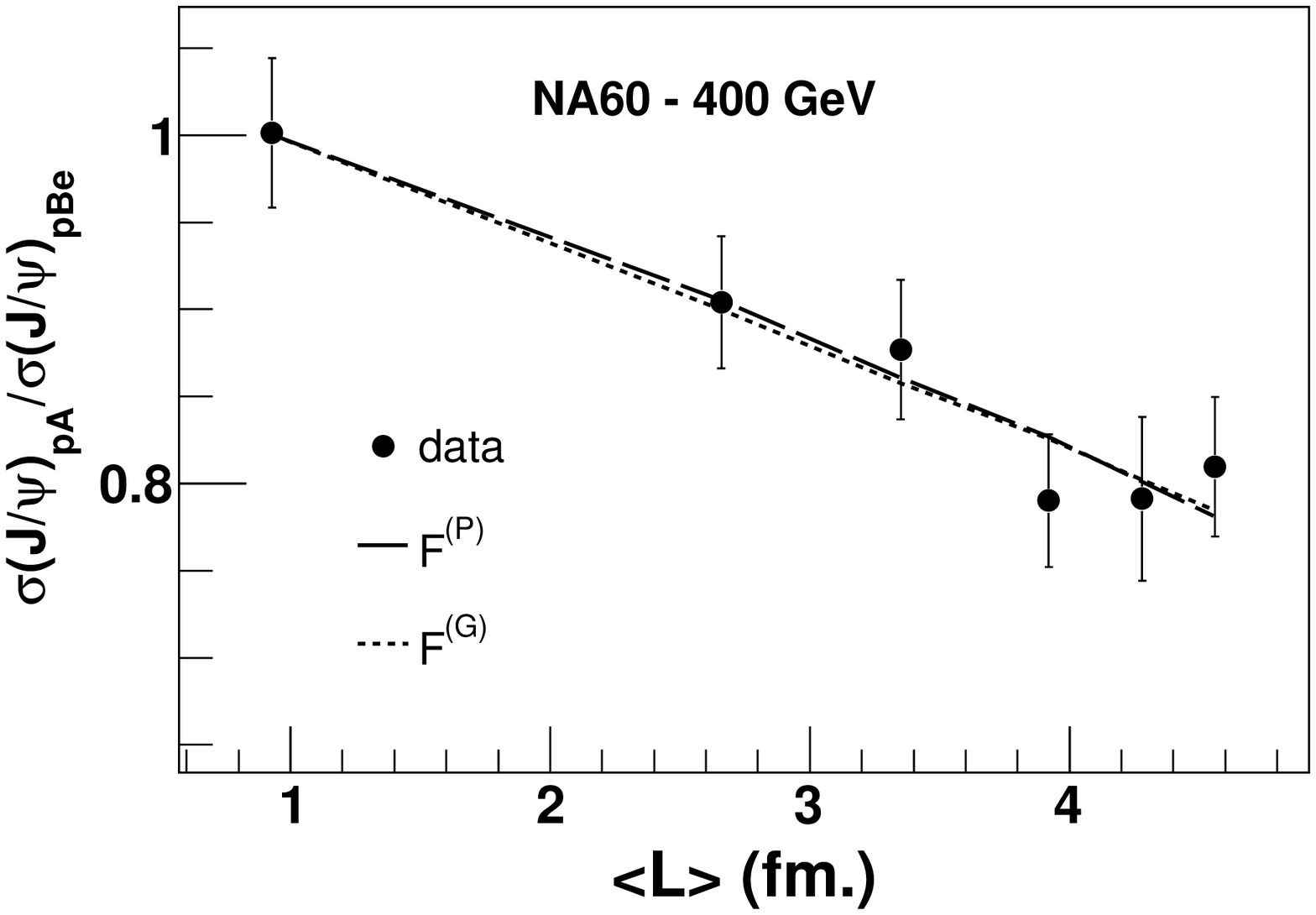}

\caption{\footnotesize J/$\psi$ cross section ratios for p-A collisions at different energies of the incident proton beam from different fixed target experiments. Data are fitted with model calculations for two different forms of the transition probability representing two different physical mechanisms of J/$\psi$ formation from an initial $c\bar{c}$ pair.}
\label{fig7}
\end{figure*}

\begin{table*}[ht]\centering
\caption{Basic features of the experiments providing the charmonium
production cross sections (or ratios) considered in the present study.}
\label{tabII}
\vglue4mm
\begin{tabular}{|c|c|c|c|} \hline

Experiment & E$_{\rm lab}$~[GeV] & Collision systems & Phase space\\ \hline
NA60~\cite{NA60} \rule{0pt}{0.5cm} & 158 & p-Be,\,Al,\,Cu, \,In, \,W, \,Pb, \,U & $0.28 < y_{\rm cms} < 0.78$ \\\hline
NA50~\cite{NA50-200} \rule{0pt}{0.5cm} & 200 & p-C,\,Al,\,Cu, \,Ag, \,Pb, \,U & $0.0 < y_{\rm cms} < 1.0$ \\\hline
NA50~\cite{NA50-400} & 400 & p-Be,\,Al,\,Cu,\,Ag,\,W,\,Pb & $-0.425 < y_{\rm cms} < 0.575$ \\\hline

NA60~\cite{NA60} & 400 & p-Be, \,Cu,\,Ag,\,W,\,Pb, \,U  & $-0.17 < y_{\rm cms} < 0.33$ \\\hline

NA50~\cite{NA50-450} & 450 & p-Be,\,Al,\,Cu,\,Ag,\,W & $-0.50 < y_{\rm cms} < 0.50$ \\\hline
E866~\cite{E866}     & 800 & p-Be,\,W & $-0.10 < x_{\rm F} < 0.93$\\\hline
\mbox{HERA-B}~\cite{HERAB} & 920 & p-C,\,W & $-0.34 < x_{\rm F} < 0.14$ \\ \hline 

\end{tabular}
\end{table*}

We now examine the J/$\psi$ production for nuclear targets. In this case the production is influenced by the interplay of variety of physical processes. For a thorough understanding of the involved mechanisms, we need to have accurate sets of data, spanning over large intervals of the incident proton beam energy and covering large $x_F$ and $p_T$ intervals. As far as present experimental situation is concerned, a wealth of high statistics J/$\psi$ data samples have been collected over the years in several fixed target experiments covering different kinematic range and energy domains. The HERA-B \cite{HERAB} experiment at DESY has recorded J/$\psi$ production in p-C, p-W reactions at 920 GeV incident proton beam energy. E866 \cite{E866}  at FNAL has studied p-Be, p-Fe and p-W collisions at 800 GeV, CERN-SPS experiment NA50 at 200 \cite{NA50-200}, 400 \cite{NA50-400} and 450 \cite{NA50-450}  GeV and NA60 \cite{NA60}  at 158 and 400 GeV for several different nuclear targets. A unanimous feature observed in all such measurements is that at fixed collision energy, charmonium production rate per target nucleon decreases with increasing A, the mass number of target nucleus. Different experimental approaches are in vogue to quantify the reduction. Usual way of parameterizing the nuclear effect is to fit the A dependence of the production cross section with simple power law: $\sigma_{pA}=\sigma_{0} \times A^{\alpha}$, and then to study the evolution of $\alpha$ with $x_F$ and $p_T$. Measured values of $\alpha$ are found to be less than unity which then indicates suppression in nuclear medium. Alternatively, the nuclear effects have also been expressed by the fitting the data in the framework of Glauber model having as input parameters the inelastic nucleon-nucleon cross section and density distributions for various nuclei. The model gives as an output $\sigma_{abs}^G$, which represents the break-up cross section of the $c\bar{c}$ pair in its pre-resonance or resonance state. However, most often  the detailed Glauber calculation, is approximated, in first order, by a simplified $\rho<L>$ exponential parameterization: $\sigma_{pA} = \sigma_{0} A exp(-\sigma_{abs}^{\rho<L>}\rho<L>)$. Here $\rho<L>$ denotes the average amount of nuclear matter crossed by the pre-formed charmonium state from its production point up to exiting from the nucleus.  In NA50, at 400 GeV data set, the absorption cross section for J/$\psi$ has been extracted to be 4.6 $\pm$ 0.6 mb from Glauber fit and 4.1 $\pm$ 0.5 from $\rho<L>$ fit, whereas a joint fit to both HI and LI data sets with the $\rho<L>$ parameterization reproduces a value 4.3 $\pm$ 0.7 mb at 450 GeV. By fitting the cross section ratios in the framework of Glauber model, NA60 has obtained $\sigma_{abs}^{J/\psi}$ = 7.6 $\pm$ 0.7(stat.) $\pm$ 0.6(syst.) mb. at 158 GeV and 4.3 $\pm$ 0.8(stat) $\pm$ 0.6(syst.) mb at 400 GeV. Recent measurements by NA60 collaboration have thus confirmed that the nuclear effects become more important when moving down the beam energy, an observation that remains valid when extending the comparisons to other existing sets of results. We should take note of the fact that both $\alpha$ and $\sigma_{abs}$ as reported by the experiments, are effective quantities, used to describe the convolution of all cold nuclear matter effects reducing the J/$\psi$ yield. But they do not allow to disentangle the different contributions (initial state shadowing, final state dissociation etc.) playing a role in this reduction.

\begin{table*}[ht]\centering
\caption{The $\varepsilon^2$ values obtained from fitting the data of different fixed target experiments with different energy of the incident proton beam. $\varepsilon^2_G$ and $\varepsilon^2_P$ corresponds to $F^{(P)} (q^2)$ and $F^{(G)} (q^2)$ respectively. The error bars correspond to fit errors.}

\label{tab3}
\vglue4mm
\begin{tabular}{|c|c|c|c|} \hline
Experiment & E$_{\rm lab}$~[GeV] &   $\varepsilon^2_G$~[GeV$^2$/fm] & $\varepsilon^2_P$~[GeV$^2$/fm]  \\ \hline
NA60 \rule{0pt}{0.5cm}& 158 & $3.52 \times 10^{-1} \pm 1.72 \times 10^{-2}$ & $2.47 \times 10^{-1} \pm 1.02 \times 10^{-2}$\\\hline

NA50 & 200 & $2.616 \times 10^{-1} \pm 8.72 \times 10^{-2}$ & $1.9 \time 10^{-1} \pm 5.67 \times 10^{-2}$\\ \hline

NA50 & 400 & $2.65 \times 10^{-1} \pm 1.71 \times 10^{-2}$ & $1.86 \times 10^{-1} \pm 1.07 \times 10^{-2}$\\ \hline

NA60 & 400 & $2.617 \times 10^{-1} \pm 2.22 \times 10^{-2}$ & $1.82 \times 10^{-1} \pm 1.37 \times 10^{-2}$\\ \hline

NA50 & 450 (HI) & $1.94 \times 10^{-1} \pm 2.46 \times 10^{-2}$ & $1.38 \times 10^{-1} \pm 1.6 \times 10^{-2}$ \\ \hline
NA50 & 450 (LI) & $1.79 \times 10^{-1} \pm 3.51 \times 10^{-2}$ & $1.3 \times 10^{-1} \pm 2.35 \times 10^{-2}$\\ \hline

E866 & 800 & $1.83 \times 10^{-1} \pm 3.28 \times 10^{-3}$  & $1.31 \times 10^{-1} \pm 2.18 \times 10^{-3}$ \\ \hline
HERA-B & 920 & $1.42 \times 10^{-1} \pm 1.19 \times 10^{-2}$ & $1.04 \times 10^{-1} \pm 8.16 \times 10^{-3}$\\ \hline

\end{tabular}
\end{table*}

In the present paper, we have analyzed the J/$\psi$ production cross section data measured in the energy range from 158 to 920 GeV. The data sets from different fixed target experiments, that have been used in the present analysis are given in  Table \ref{tabII} along with their phase space coverage. Among the listed experiments, NA50 presented their results as J/$\psi$ production cross section in di-muon channel for different nuclear targets as a function of the average nuclear path length $<L>$. On the other hand NA60 has presented their results as the ratio of J/$\psi$ production cross sections for a particular nucleus to that due to Be as a function of  $<L>$. In case of any model study, one advantage to work with ratios is that we can get rid of the multiplicative parameters as they get canceled. Hence for NA50 as well, we have fitted the model results with the cross section ratios rather than their absolute values. For 400 and 450 GeV we have used the lightest available target Be as the reference. In case of 200 GeV,  we have chosen C as our reference instead of H, as the nuclear suppression are found to be sensitive to the choice of the lightest target \cite{Lourenco}. Fig.~\ref{fig7} below illustrates our calculations fitted with the corresponding data sets from NA50 and NA60. On the other hand, at 800 (920) GeV we have fitted the ratio of J/$\psi$ production cross section for W to C (Be) as a function of $x_F$, calculated from the parameter $\alpha$ and given in \cite{Lourenco}. $L(A)$, as introduced in Eq.~(\ref{q2shift}), is replaced by the respective value $<L>$ for a target nucleus, as used in the corresponding data set at a particular fixed target energy. The value of $\varepsilon^2$ (the relevant parameter in the model for quantifying the final state nuclear dissociation of J/$\psi$) at each energy is extracted by fitting the available data independently with the model results for both parameterizations of the transition probability namely the Gaussian distribution $F^{(G)} (q^2)$ (simulating the color singlet model) and the power law distribution $F^{(P)} (q^2)$ (mimicking the color octet formalism). The transition probability $F^{(C)} (q^2)$ containing the essential features of color evaporation model (CEM) \cite {Evap}, is a special case of $F^{(P)} (q^2)$ for $\alpha_F = 0$. It assumes that all the $c\bar{c}$ pairs with invariant mass less than open charm threshold has the same constant probability to evolve in a J/$\psi$ meson. However as pointed out in \cite{Qui}, the pairs just below the threshold should have a smaller probability to become J/$\psi$, than those far below the threshold. In other words, the power law parameterization $F^{(P)} (q^2)$, with $\alpha_F$ $>$ 0 is believed represent more accurate physics for J/$\psi$ production. We thus refrain ourselves from further use of $F^{(C)} (q^2)$, while analyzing the J/$\psi$ production in p-A reactions. The best fit values of $\varepsilon^2$ as obtained separately for $F^{(G)} (q^2)$ and $F^{(P)} (q^2)$ are presented in Table \ref{tab3}. Both $\varepsilon^2_G$ and $\varepsilon^2_P$ show a non-negligible energy dependence with $\varepsilon^2_{P(G)}$ increasing with decrease in the collision energy. The observed energy dependence pattern of  $\varepsilon^2$ (characterizing the final state dissociation) is thus in line with the earlier observations of energy dependence of the J/$\psi$ absorption cross section in normal nuclear matter \cite{Lourenco}.

\begin{table*}[ht]\centering
\caption{Comparison of the extracted $\varepsilon^2$ and the corresponding absorption cross section ($\sigma_{abs}$)  values with free proton (p) and nuclear (n) parton distribution functions following EPS09 interface at different beam energies in case of Gaussian transition probability. The absorption cross sections extracted from the collected data from exponential $\rho<l>$ fitting are also included in the last column. The normal nuclear matter density is taken as $\rho_{0}= 0.15 /fm^{3}$.}
\label{tabIV}
\vglue4mm
\begin{tabular}{|c|c|c|c|c|c|c|} \hline
Experiment & E$_{\rm lab}$~[GeV] &   $\varepsilon^2_p$~[GeV$^2$/fm] & $\varepsilon^2_n$~[GeV$^2$/fm]  & $\sigma_{abs}^p$ ~[mb]   & $\sigma_{abs}^n$~[mb] & $\sigma_{abs}^{\rho<L>}$~[mb]\\ \hline
NA60 & 158 & $2.96 \times 10^{-1}\pm 1.72 \times 10^{-2}$ & $3.5\times10^{-1} \pm 1.72 \times 10^{-2}$ & $6.85 \pm 0.4$ & $8.1 \pm 0.4$ & $7.36 \pm 0.7 $ \\ \hline
NA50 & 200 & $2.2 \times 10^{-1}\pm 9.4 \times 10^{-2}$ & $2.62\times10^{-1} \pm 8.72 \times 10^{-2}$ & $5.09 \pm 2.08$ & $6.06 \pm 2.01$ & $4.68 \pm 3.55$ \\ \hline
NA50 & 400 & $1.95 \times 10^{-1}\pm 1.7 \times 10^{-2}$ & $2.65\times10^{-1} \pm 1.7 \times 10^{-2}$ & $4.51 \pm 0.39$ & $6.13 \pm 0.39$ & $4.69 \pm 0.75$  \\ \hline
NA60 & 400 & $1.87 \times 10^{-1}\pm 2.2 \times 10^{-2}$ & $2.62\times10^{-1} \pm 2.2 \times 10^{-2}$ & $4.3 \pm 0.51$ & $6.06 \pm 0.51$ & $4.44 \pm 1.02$ \\ \hline
NA50 & 450 (HI) & $1.5 \times 10^{-1}\pm 2.6 \times 10^{-2}$ & $1.94\times10^{-1} \pm 2.46 \times 10^{-2}$ & $3.47 \pm 0.6$ & $4.49 \pm 0.57$ & $3.8 \pm 1.1$ \\ \hline
NA50 & 450 (LI) & $1.34 \times 10^{-1}\pm 3.8 \times 10^{-2}$ & $1.79\times10^{-1} \pm 3.51 \times 10^{-2}$ & $3.1 \pm 0.88$ & $4.14 \pm 0.81$ & $3.71 \pm 1.51$ \\ \hline 

\end{tabular}
\end{table*}

It might be interesting to note that for the parameterization $F^{\rm (G)}(q^2)$ in Eq.~(\ref{gauss}), a shift of $q^2$ to $\bar{q}^2$ in Eq.~(\ref{q2shift}) for the J/$\psi$ suppression in nucleon-nucleus collisions gives the following relation 
\begin{equation}
\sigma_{pA\rightarrow {\rm J/}\psi}(\sqrt s) =
\exp\left[-\frac{\varepsilon^2}{2\alpha_F^2}\, L(A)\right]
\sigma_{NN\rightarrow {\rm J/}\psi}(\sqrt s).
\label{ABgauss}
\end{equation}

This relation is effectively the same as that accounted by the exponential $\rho<L>$ parameterization. One can thus calculate the corresponding absorption cross section as $\sigma_{\rm abs} (mb) = (10 \times \varepsilon^2)/(2 \times \alpha_F^2 \times \rho_0)$. Using the model parameters at 158 GeV, and taking $\rho_{0}\sim 0.15/fm^{3}$, the corresponding absorption cross section comes out to be $\sigma_{\rm abs}\sim 8.15 \pm 0.4$~mb. Though close, the value is still larger than that reported by NA60 (with a full Glauber model analysis). As pointed out earlier, the $\sigma_{\rm abs}$ as reported by different experimental groups are the effective quantities since they include both initial as well as final state effects. However in our calculation $\varepsilon^2$ and hence the corresponding $\sigma_{\rm abs}$ characterizes only the size of the final state dissociation. We can calculate the effective $\varepsilon^2$ (for $F^{(G)}(q^2)$) at different energies, following our adopted model, if we neglect the initial state effect in our calculation and use free proton pdfs without any nuclear modification for calculation of $c\bar{c}$ pair production cross section. Table \ref{tabIV} shows a comparison between the best fitted $\varepsilon^2$ values at different SPS energies with and without the nuclear modification of the parton distribution functions and along with the corresponding $\sigma_{\rm abs}$. In addition, the effective absorption cross sections extracted from the exponential fitting of the experimentally measured data points are also shown. 
 
\begin{figure} \vspace{-0.1truein}
\includegraphics[width=8.5cm]{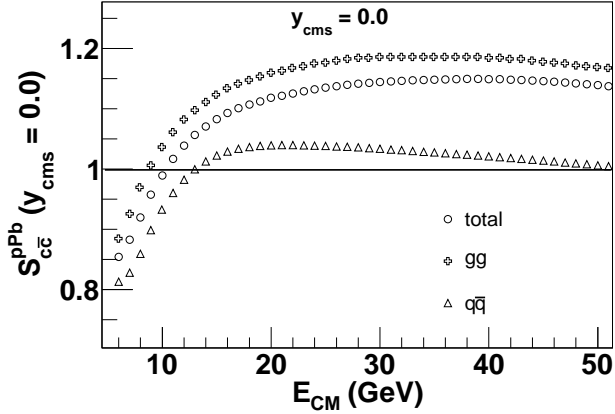}
\caption{\footnotesize Energy evolution of the shadowing factor $S^{pPb}_{c\bar{c}}$ at mid-rapidity, following EPS09 parameterization. The individual contribution due to light quark annihilation (open triangles) and gluon fusion (open crosses) are shown separately along with their joint contribution (open circles).}
\label{fig8}
\end{figure}

\begin{figure} \vspace{-0.1truein}
\includegraphics[width=8.5cm]{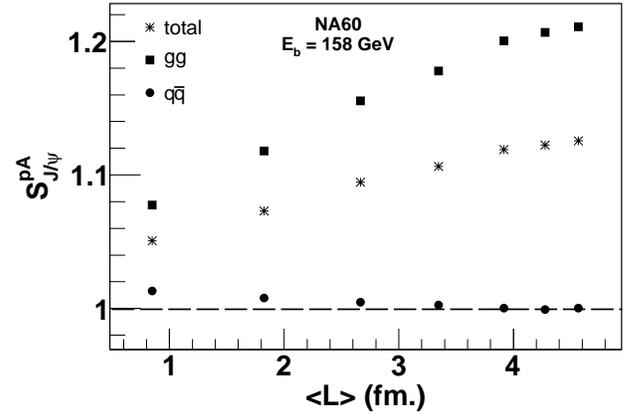}
\includegraphics[width=8.5cm]{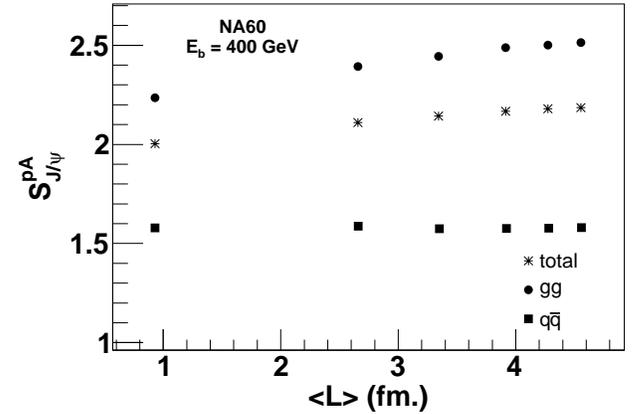}
\caption{\footnotesize Variation of the shadowing factor $S^{pA}_{J/\psi}$ as function of $<L>$ at NA60 for two different energies $E_{b}$ = 158 GeV (upper panel) and $E_{b}$ = 400 GeV (lower panel) of the incident proton beam. The contribution of the light quark annihilation and gluon fusion are separately shown along their sum. The inclusive production cross sections are obtained by integrating the differential cross sections over the respective rapidity windows corresponding to the two beam energies. The Gaussian form of transition probability is used for evolution of $c\bar{c}$ pairs into J/$\psi$ mesons.}
\label{fig9}
\end{figure}

It appears from the Table \ref{tabIV} that if we incorporate, as initial state effect, the nuclear modification of the parton densities following EPS09 interface, we need a higher value of $\varepsilon^2$ than that of a free proton pdf to reproduce the corresponding data, in the SPS energy regime. This can be understood by studying the influence of the nuclear modification of the parton distribution functions to the J/$\psi$  production. For this purpose let us define the shadowing factor ($S^{pA}_{J/\psi}$) for a particular nucleus as the ratio between the J/$\psi$ production cross section (per nucleon-nucleon) collision in p-A and p-p:

\begin{equation}
S_{pA}^{\rm J/\psi}(y_{cms})=\frac{1}{A}\frac{d\sigma_{\rm J/\psi}^{pA}/dy_{cms}}{d\sigma_{\rm
J/\psi}^{pp}/dy_{cms}}
\label{eq:4}
\end{equation}

Thus $S^{pA}_{J/\psi}$ less than unity indicates shadowing whereas $S^{pA}_{J/\psi}$ larger than unity signifies anti-shadowing. Since the pdfs enter as an input to the calculation of perturbative $c\bar{c}$ pair production cross section, hence in Fig.\ref{fig8}, we plot $S^{pPb}_{c\bar{c}}$ at mid-rapidity, as a function of center of mass energy, in the domain of different fixed target experiments. In the energy regime probed by the SPS experiments, charm production at mid-rapidity explores x (fraction of nucleon momentum carried by partons) values corresponding to the anti-shadowing region, where parton densities in the nuclei are enhanced with respect to those of free protons. In the absence of any final state interaction this would certainly result in enhancement of J/$\psi$ production cross section per nucleon in p-Pb collisions compared to p-p collisions. However in the lower energy regime, relevant for FAIR experiments, EPS09 parameterization shows a shadowing effect resulting in the reduction of the produced $c\bar{c}$ pairs in p-Pb collisions compared to the p-p collisions at a given energy. 

\begin{figure} \vspace{-0.1truein}
\includegraphics[width=8.5cm]{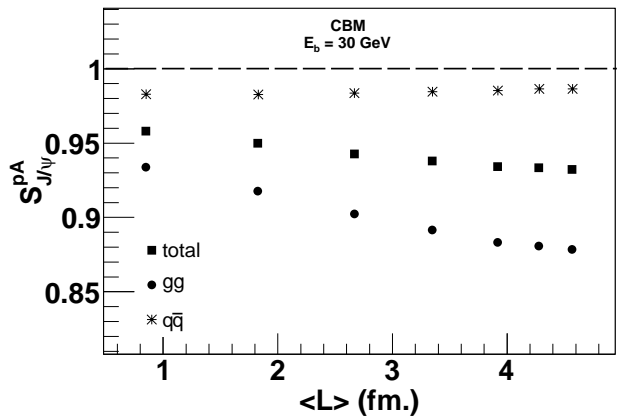}
\caption{\footnotesize Variation of the shadowing factor $S^{pA}_{J/\psi}$ as function of $<L>$ for a 30 GeV incident proton beam. Such measurements will be performed in the SIS-100 phase of the upcoming  CBM experiment at FAIR.}
\label{fig10}
\end{figure}
To make our study more quantitative let us now calculate $S^{pA}_{J/\psi}$ for different nuclei as a function of $<L>$ at NA60 beam energies, corresponding to their respective kinematic interval. The results are depicted in Fig. (\ref{fig9}). As it appears from the figure, at both the energies we obtain an over all anti-shadowing for all the nuclei, resulting in an increment in the J/$\psi$ production cross section compared to the p-p reactions in absence of any final state dissociation. At 158 GeV, the $q\bar{q}$ fraction of the cross section shows a very weak shadowing for heavier nuclei, which is more than counterbalanced by the stronger contribution from gluon fusion. Hence if nuclear modification of parton densities is taken into account, one needs a more severe amount of final state multiple scattering, than with a free proton pdf, to reproduce the measured data collected at SPS, which actually show a suppression in yield as a function of $<L>$. Naturally the absorption cross section extracted from the corresponding $\varepsilon^2$ will also be larger if one takes into account the nuclear pdfs following EPS09 interface compared to the free proton (as well as the experimentally measured) case. The observation is in agreement with the earlier studies \cite{NA60, Lourenco}. The results are sensitive to the adopted parameterization of the nuclear parton densities. For example, the NA60 collaboration has observed that if the initial state effects are evaluated with EKS98 \cite{EKS98} parameterization, a larger absorption cross section is required to fit the measured data ($\sigma_{abs}^{J/\psi}$ (158 GeV) = 9.3 $\pm 07 \pm 0.7$ mb and $\sigma_{abs}^{J/\psi}$ (400 GeV) = 6.0 $\pm$ 0.9 $\pm$ 0.7 mb). Slightly higher $\sigma_{abs}^{J/\psi}$  has been obtained using EPS08 \cite{EPS08} parameterization. However the situation might be different at FAIR as seen from Fig.~\ref{fig10}. The inclusive production cross sections are calculated over a rapidity interval $-0.5 \le y_{cms} \le 0.5$. Incorporation of nuclear pdfs following EPS09 parameterization, results in reduction of the $c\bar{c}$ and hence J/$\psi$ production in nuclear targets compared to the protonic collisions. These observations can be understood by looking at the nature of EPS09 parameterization in Fig. \ref{fig2}. The rapidity coverage of NA60 at 158 GeV is $0.28 < y_{cms} < 0.78$. This roughly corresponds to a $x_b (= (m_{J/\psi}/\sqrt{s})exp(-y_{cms}))$ regime, $\sim 0.08 < x_{b} < 0.13$, where we observe anti-shadowing. However the rapidity interval at 30 GeV beam energy considered in the present calculation explores a $x_b$ region, $\sim 0.24 < x_b < 0.66$. In this region parton population in a nucleon bound inside a nucleus is depleted compared to a free nucleon. Thus, within the present theoretical scenario, a smaller dissociation cross section, compared to the free proton case might be required to describe the J/$\psi$ suppression expected to occur at FAIR. However a conclusive picture can be drawn only after the measurements are performed.

\begin{figure} \vspace{-0.1truein}
\includegraphics[width=8.5cm]{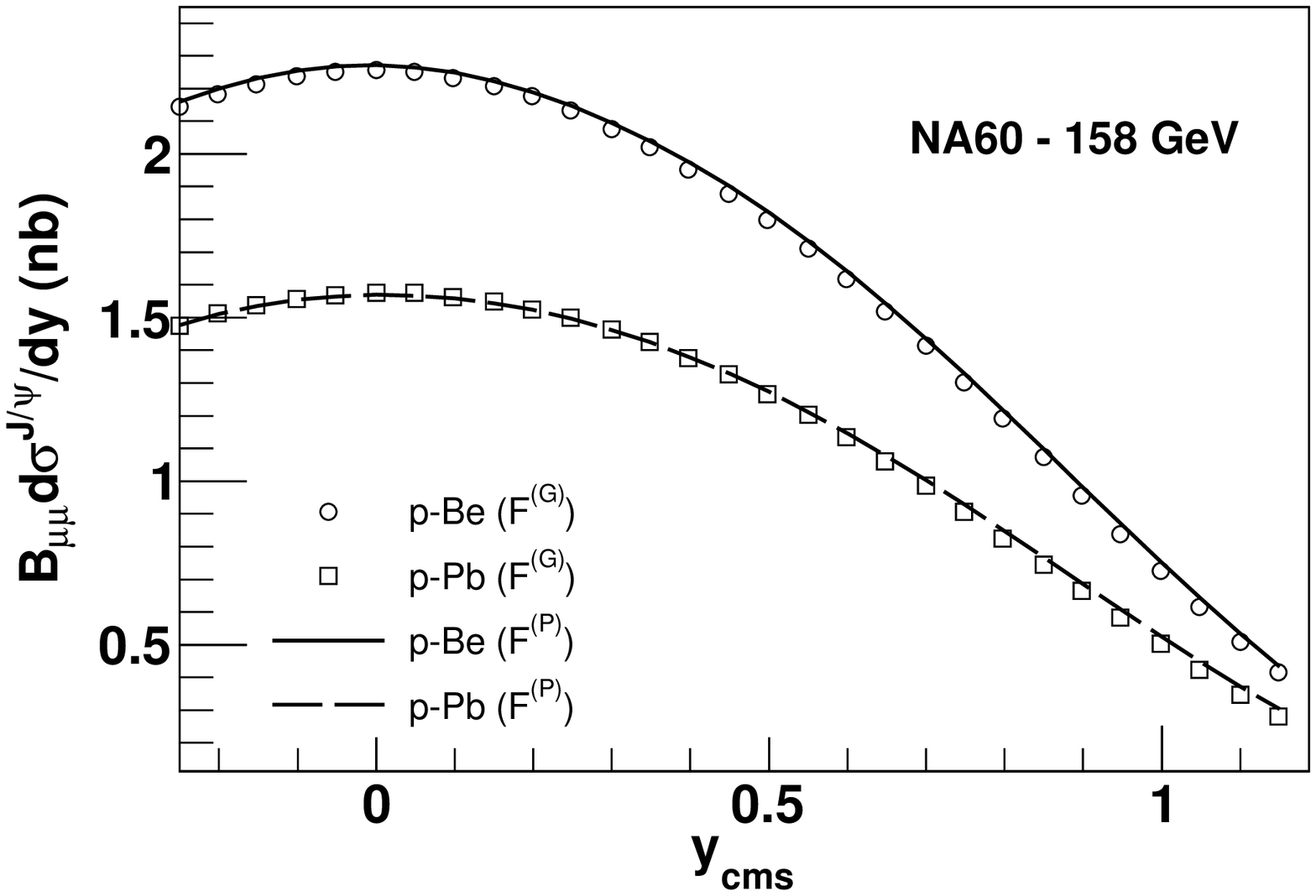}
\includegraphics[width=8.5cm]{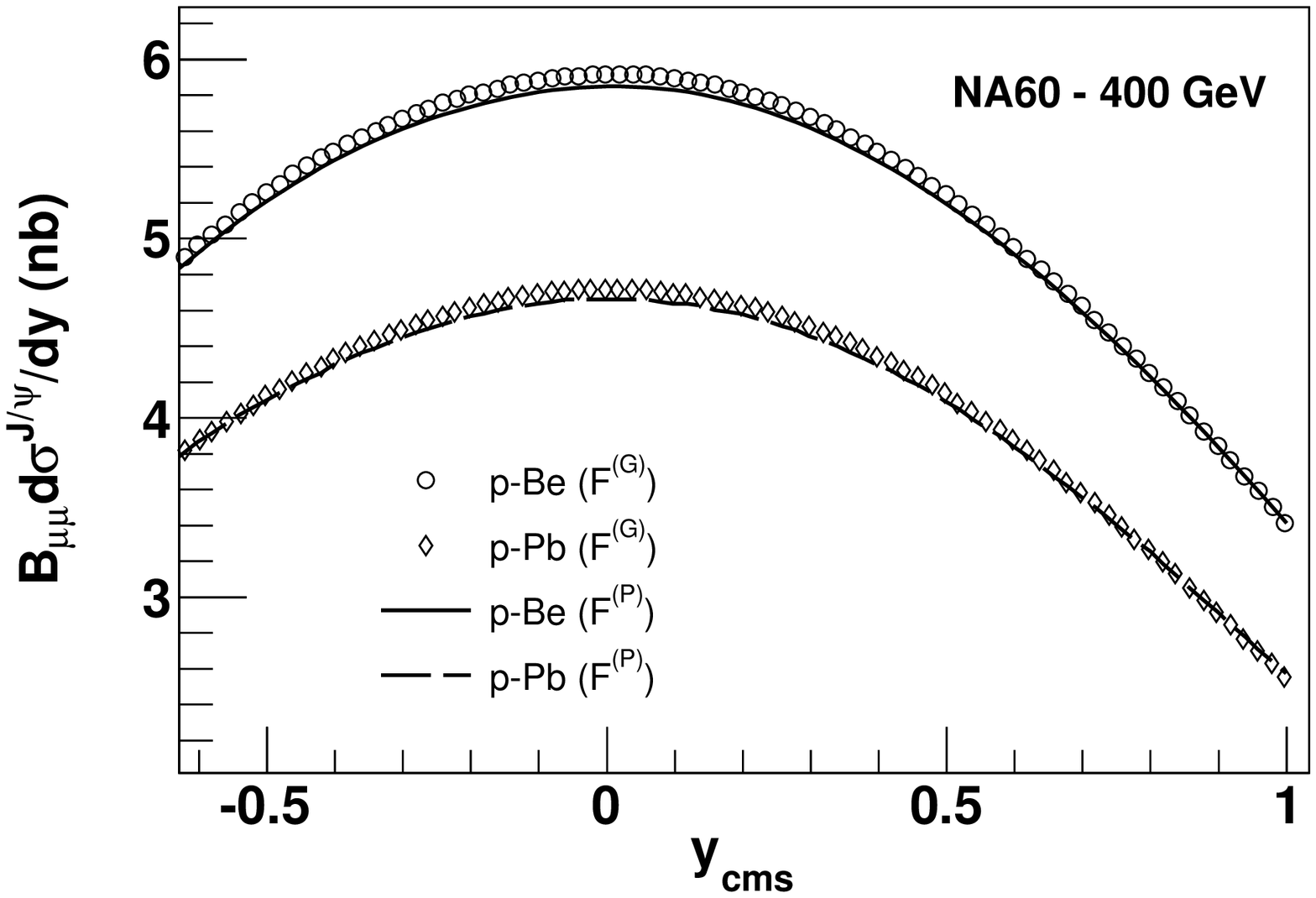}
\caption{\footnotesize J/$\psi$ rapidity distributions in p-A collisions for Be and Pb targets at 158 GeV (top) and 400 GeV (bottom) as obtained in our calculations. Two different parametric forms of the transition probabilities produce almost similar results.}
\label{fig11}
\end{figure}

Apart from the inclusive production cross sections, NA60 has also measured the J/$\psi$ rapidity distributions \cite{NA60} for several target nuclei at 158 GeV as well as 400 GeV. Within the experimental coverage of NA60 di-muon spectrometer, the rapidity distributions are found to be well reproduced by Gaussian functions. Preliminary analysis reports that at 158 GeV, the distributions corresponding to the different p-A collisions can simultaneously be described, by a single Gaussian function, having mean ($\mu_{y}$) = 0.05 $\pm$ 0.05 and sigma ($\sigma$) = 0.51 $\pm$ 0.02. At 400 GeV the corresponding fit parameters are mean ($\mu_{y}$) $\sim$ -0.2 (guided by the previous NA50 measurements at the same energy) and ($\sigma$) = 0.81 $\pm$ 0.03. However the data for rapidity distribution are presented in arbitrary units. Cross section values have not yet been published. We thus compare the shape of rapidity distributions at both energies, as obtained from our calculations with the experimentally measured ones. The rapidity distributions for different collision systems are simultaneously fitted by a single three parameter Gaussian function. For an unbiased comparison we have also performed the Gaussian fitting of the experimental data points. Results are given in Table \ref{tabIIIa}. At both the energies, the widths of the rapidity distribution are found to be higher in our model calculations in comparison with the measured values. Moreover our calculations can not reproduce the negative mean value of the rapidity distribution at 400 GeV. For illustration, we have shown the rapidity distributions for Be and Pb targets, following our model calculations in Fig. \ref{fig11}.

\begin{table*}[ht]\centering
\caption{Parameters of the Gaussian functions used to fit the J/$\psi$ rapidity distributions in p-A collisions at 158 GeV and 400 GeV. The superscripts (P) and (G) respectively denote the power law and Gaussian parameterizations.}
\label{tabIIIa}
\vglue4mm
\begin{tabular}{|c|c|c|c|c|c|c|} \hline
E$_{Lab}$~[GeV] & $\mu_y^{(Expt)}$ & $\mu_y^{(P)}$ & $\mu_y^{(G)}$ & $\sigma_y^{(Expt)}$ & $\sigma_y^{(P)}$ & $\sigma_y^{(G)}$ \\ \hline
158 & $0.044\pm0.1$ & $0.02\pm0.087$ & $0.02\pm0.086$ & $0.55\pm0.06$ & $0.66\pm0.09$ & $0.065\pm0.096$ \\ \hline
400 & $-0.27\pm0.16$ & $9.17\times10^{-4}\pm0.037$ & $-1.93\times10^{-4}\pm0.034$ & $0.86\pm0.2$ & $0.97\pm0.07$ & $0.92\pm0.059$ \\ \hline

\end{tabular}
\end{table*}

\section{J/$\psi$ PRODUCTION AT FAIR}

\begin{figure} \vspace{-0.1truein}
\includegraphics[width=8.5cm]{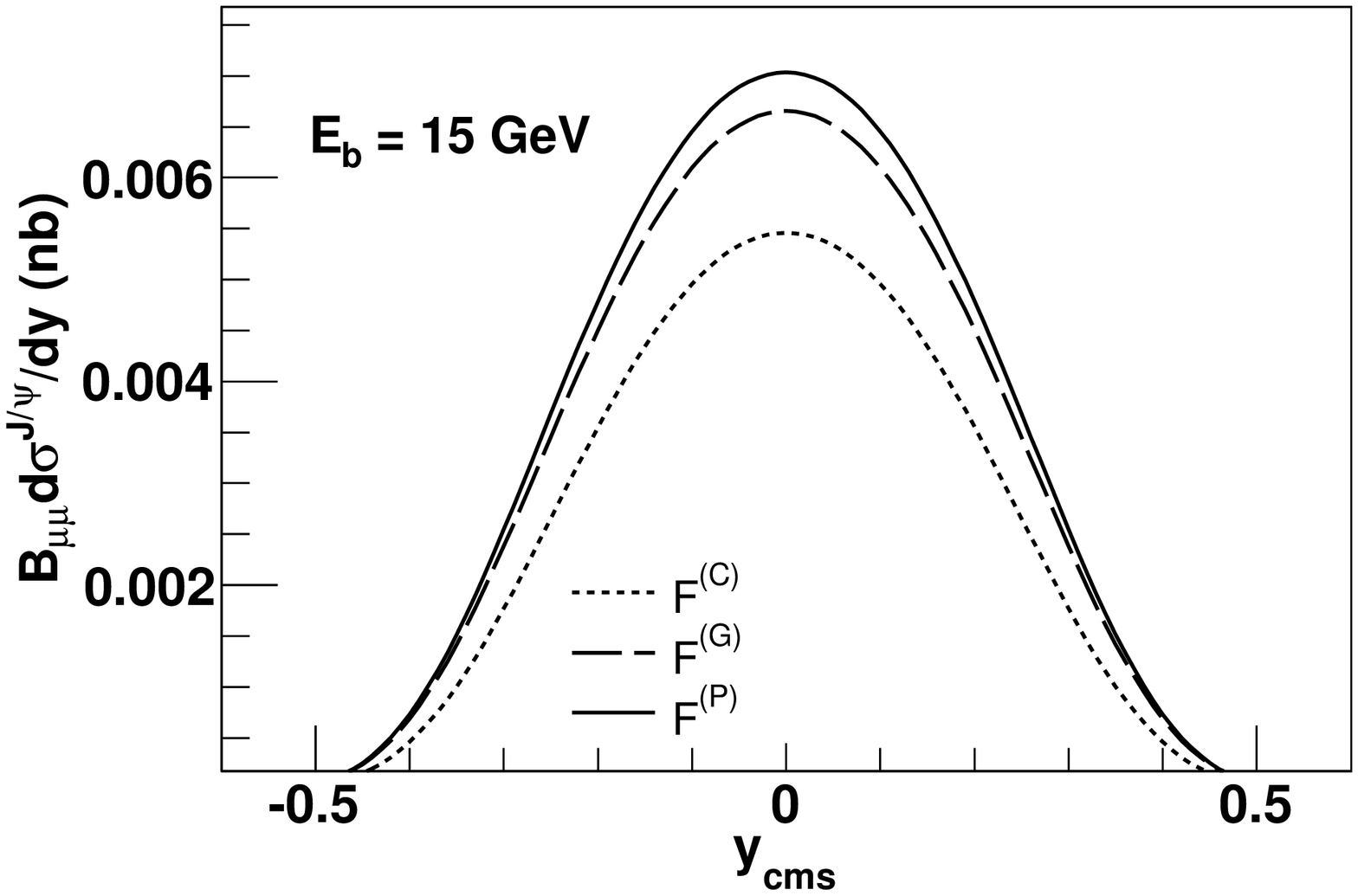}
\includegraphics[width=8.5cm]{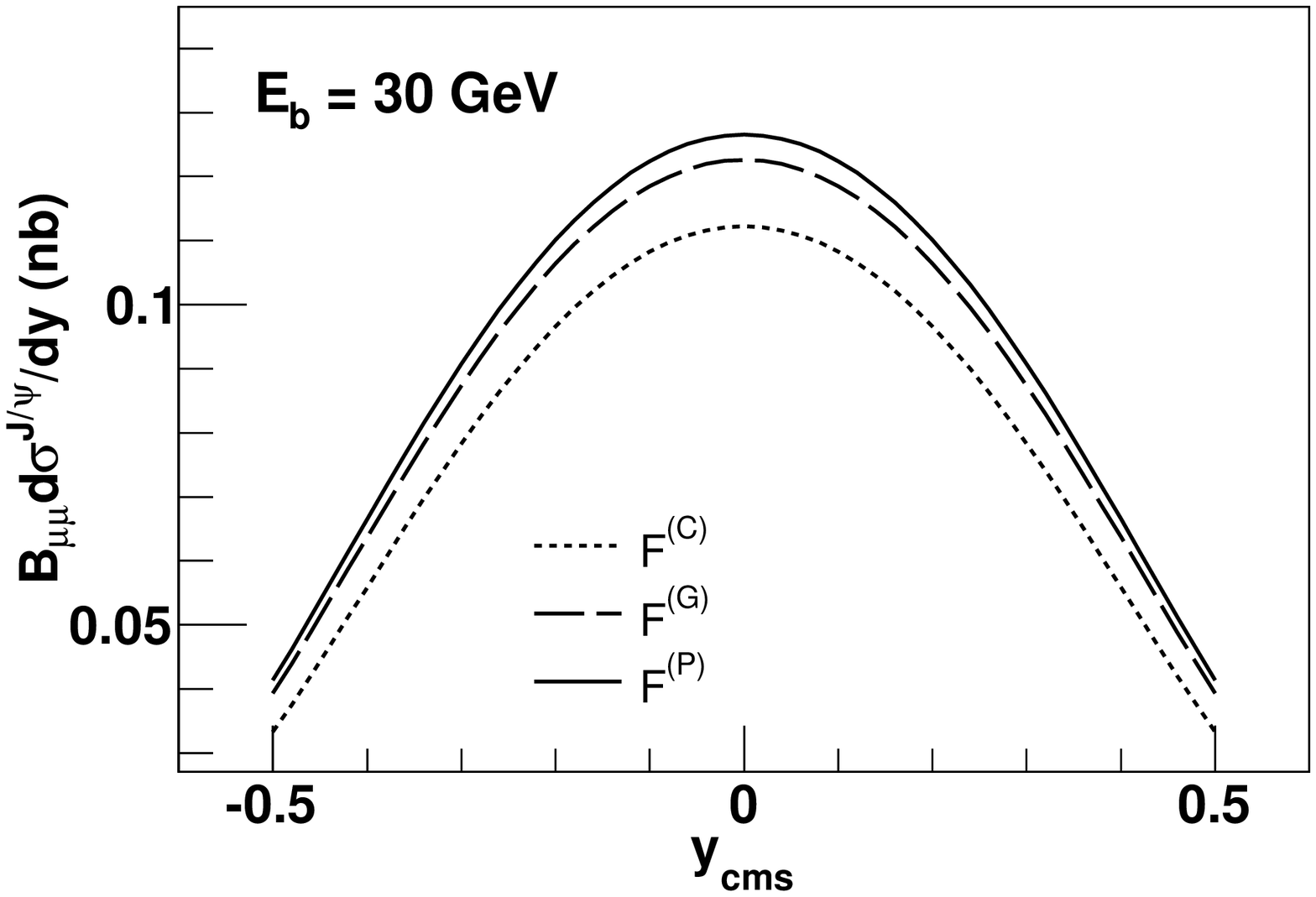}
\includegraphics[width=8.5cm]{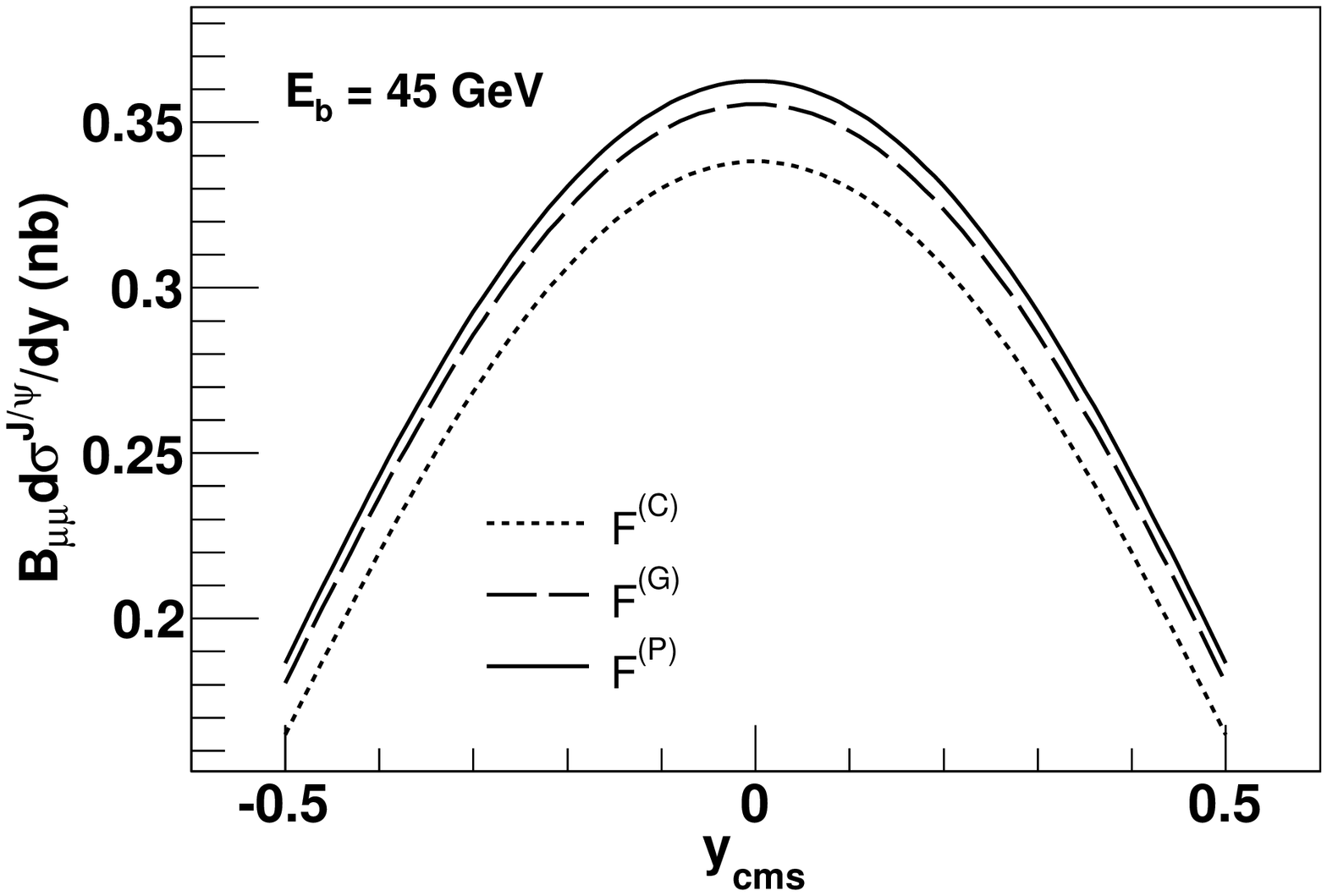}
\caption{\footnotesize Rapidity distribution of J/$\psi$ mesons produced in pp collisions in di-muon channel, for 15 GeV (upper panel), 30 GeV (middle panel) and 45 GeV (bottom panel) incident proton beams. Such measurements will be performed in the SIS-100 and SIS-300 rings of the upcoming  CBM experiment at FAIR.}
\label{fig12}
\end{figure}

Our ultimate goal is to estimate the J/$\psi$ yield in p-p and p-A collisions at energy values planned to be explored at FAIR. In the start version of FAIR (SIS-100), proton beams up to 30 GeV will be accelerated. In the final version high intensity proton beams with energy up to 45 GeV will be available. In Fig.~\ref{fig12}, we present the J/$\psi$ rapidity distribution in protonic collisions with 15, 30 and 45 GeV proton beams, for all three different functional forms of the transition probability. As expected, the distributions are symmetric about the center of mass rapidity $y_{cms}$ = 0, and the width of the distribution decreases with the decrease in the beam energy. The data for inclusive J/$\psi$ production cross section, in this energy regime, are well accounted by all three different cases of formation mechanisms as seen in the previous section. However the differential cross sections take different values for different form of $F_{c\bar{c}\rightarrow {\rm J/}\psi}(q^2)$. The yield is found to be minimum for $F^{(C)} (q^2)$ and maximum for $F^{(P)} (q^2)$ for all three beam energies, with the relative difference decreasing with increasing beam energy.  
   
\begin{figure} \vspace{-0.1truein}
\includegraphics[width=8.5cm]{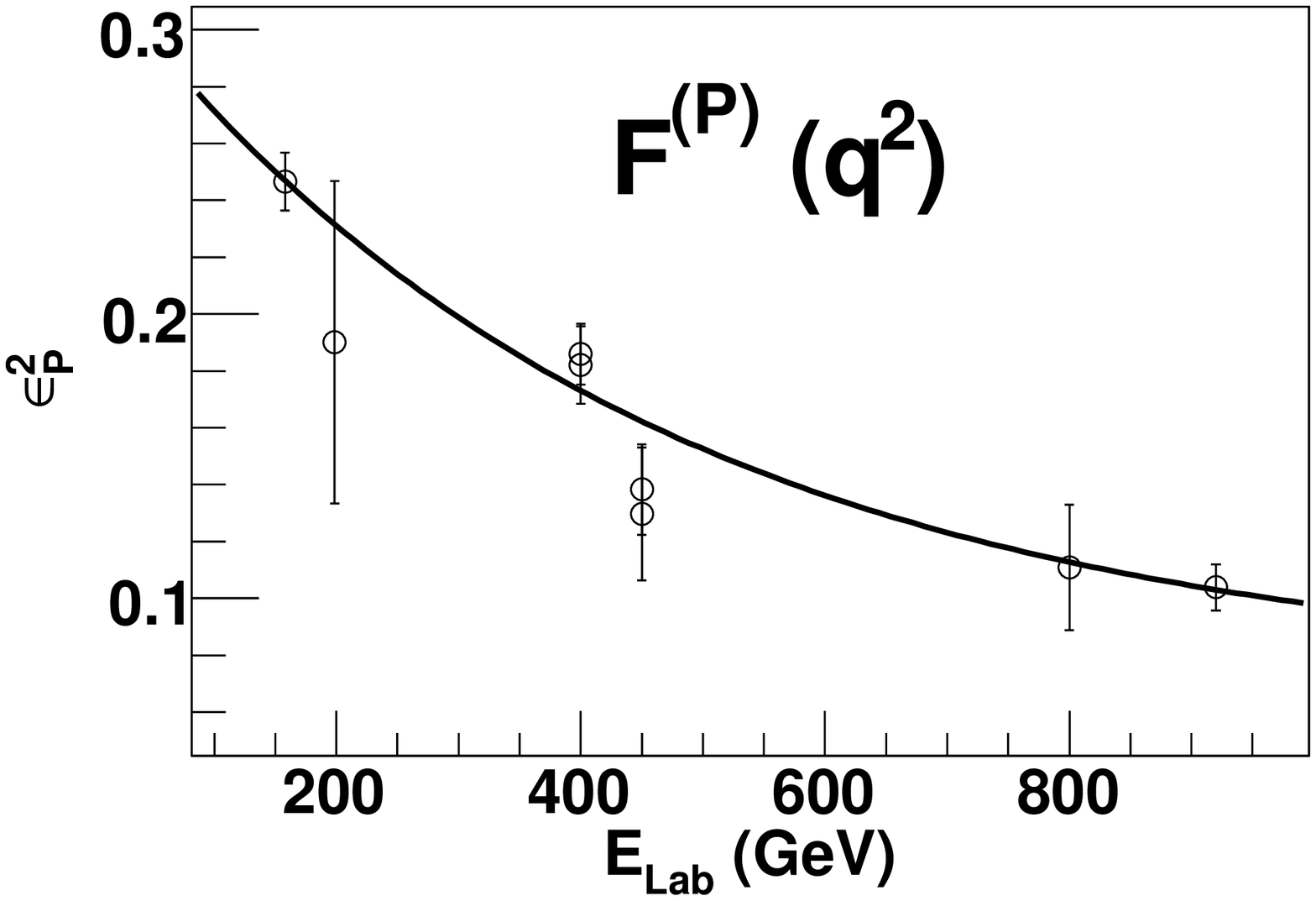}
\includegraphics[width=8.5cm]{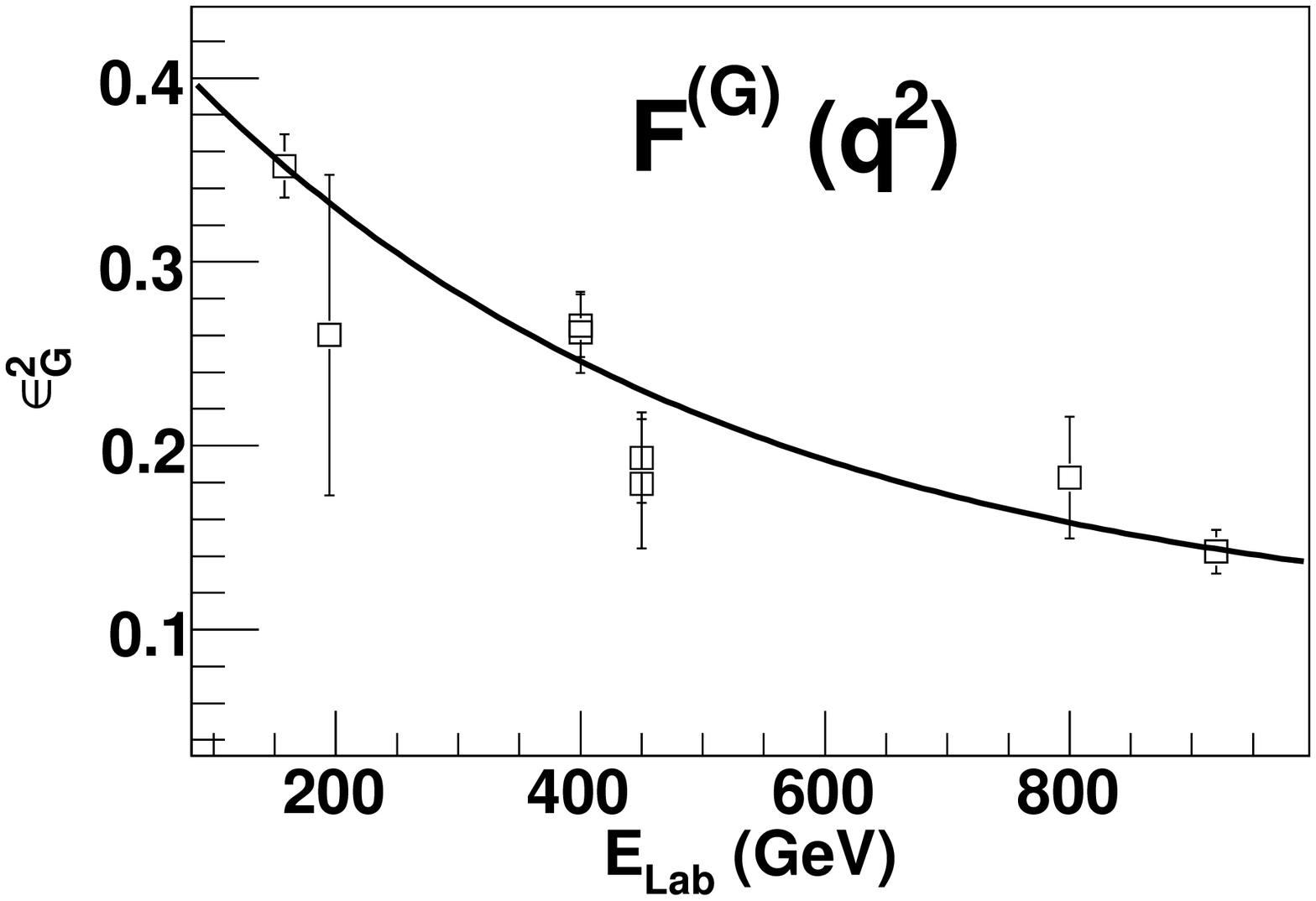}
\caption{\footnotesize Variation of $\varepsilon^2$ with energy of the incident proton beam for a) $F^{(P)}(q^2)$ (top panel) and b) $F^{(G)}(q^2)$ (bottom panel). The points on the plots are taken from \ref{tab3}. In both the cases the energy dependence has been parameterized following exponential functions.}
\label{fig13}
\end{figure}

\begin{figure*} \vspace{-0.1truein}
\includegraphics[width=8.5cm]{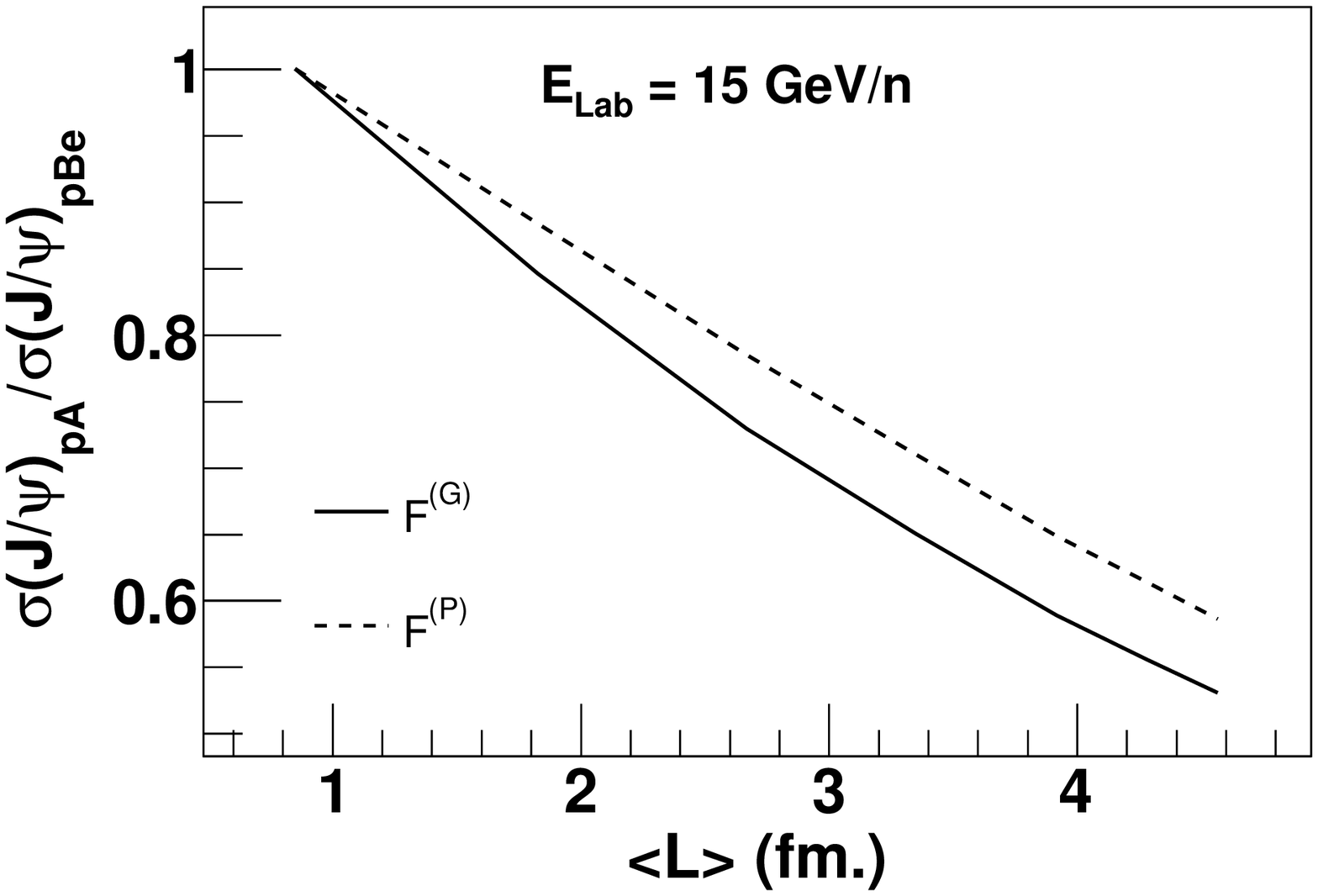}
\includegraphics[width=8.5cm]{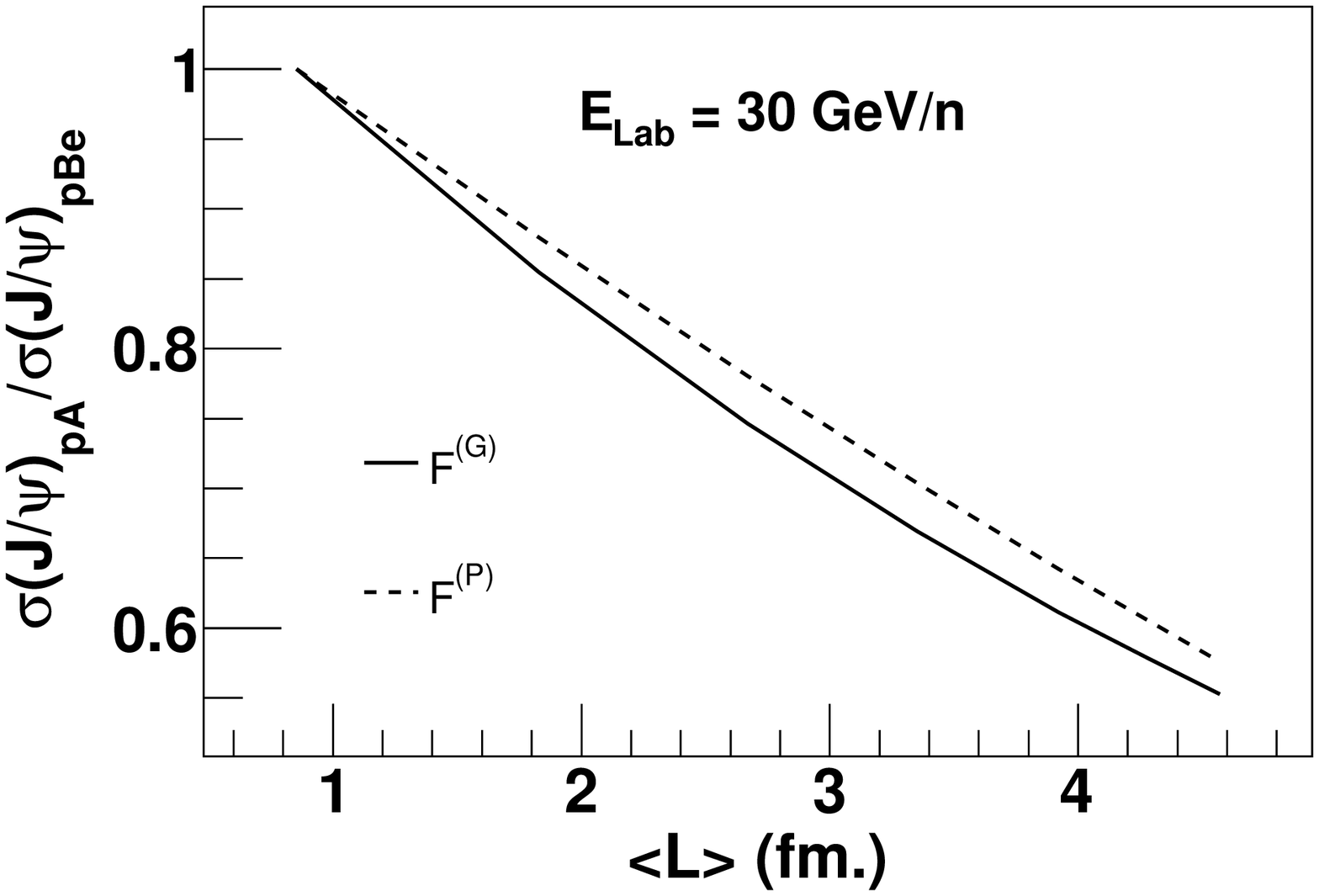}
\includegraphics[width=8.5cm]{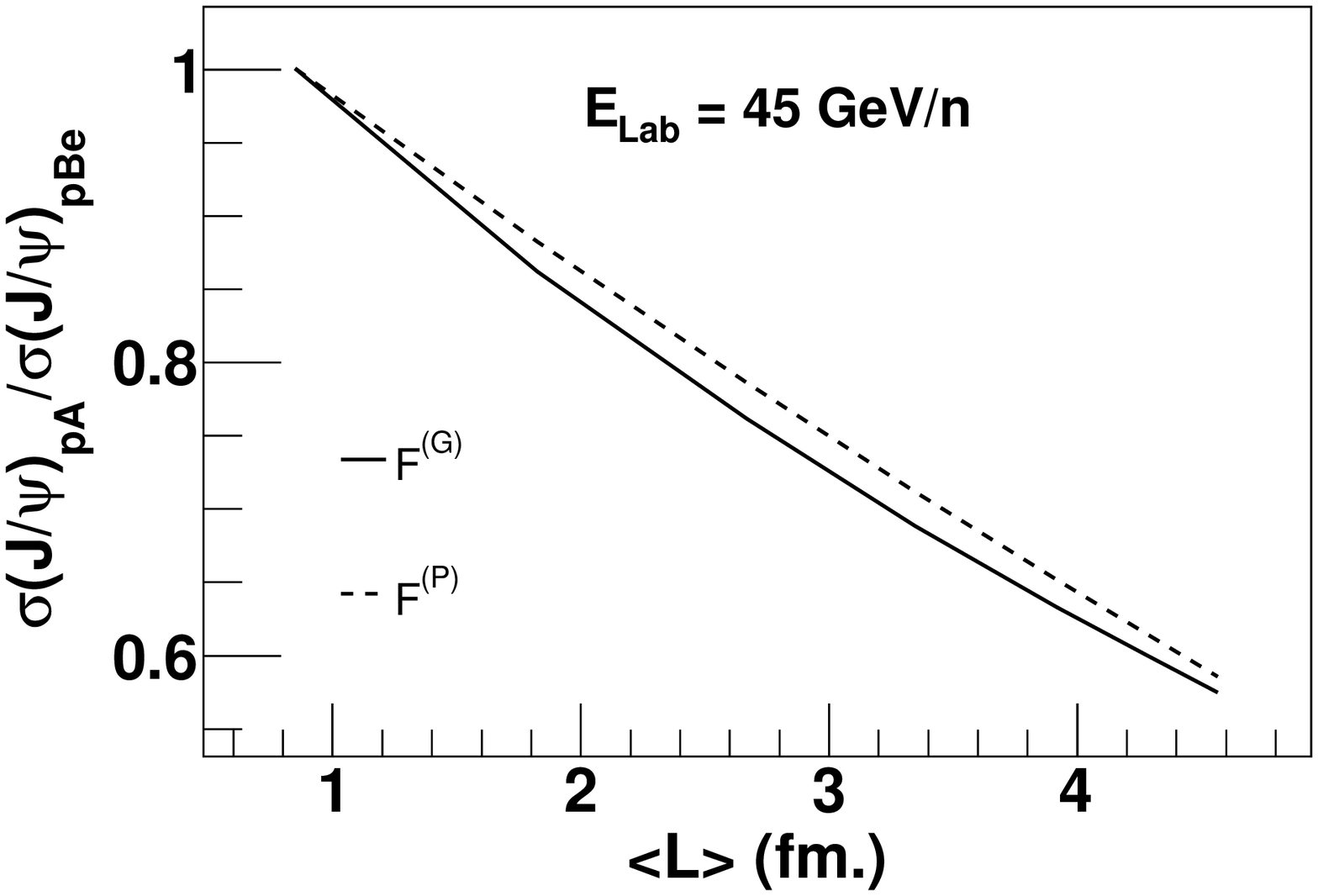}

\caption{\footnotesize J/$\psi$ production cross section ratios as a function of $<L>$ for different target nucleus in the FAIR energy domain, as predicted by our calculations. We have used both the Gaussian $F^{(G)}(q^2)$ as well as power law $F^{(P)}(q^2)$ forms of the transition probabilities for calculation of J/$\psi$ yield. Lower be the energy of the incident proton beam the two different forms representing two different physical mechanisms of J/$\psi$ formation is seen to generate different amount of suppression.}
\label{fig14}
\end{figure*}

\begin{figure*} \vspace{-0.1truein}
\includegraphics[width=8.5cm]{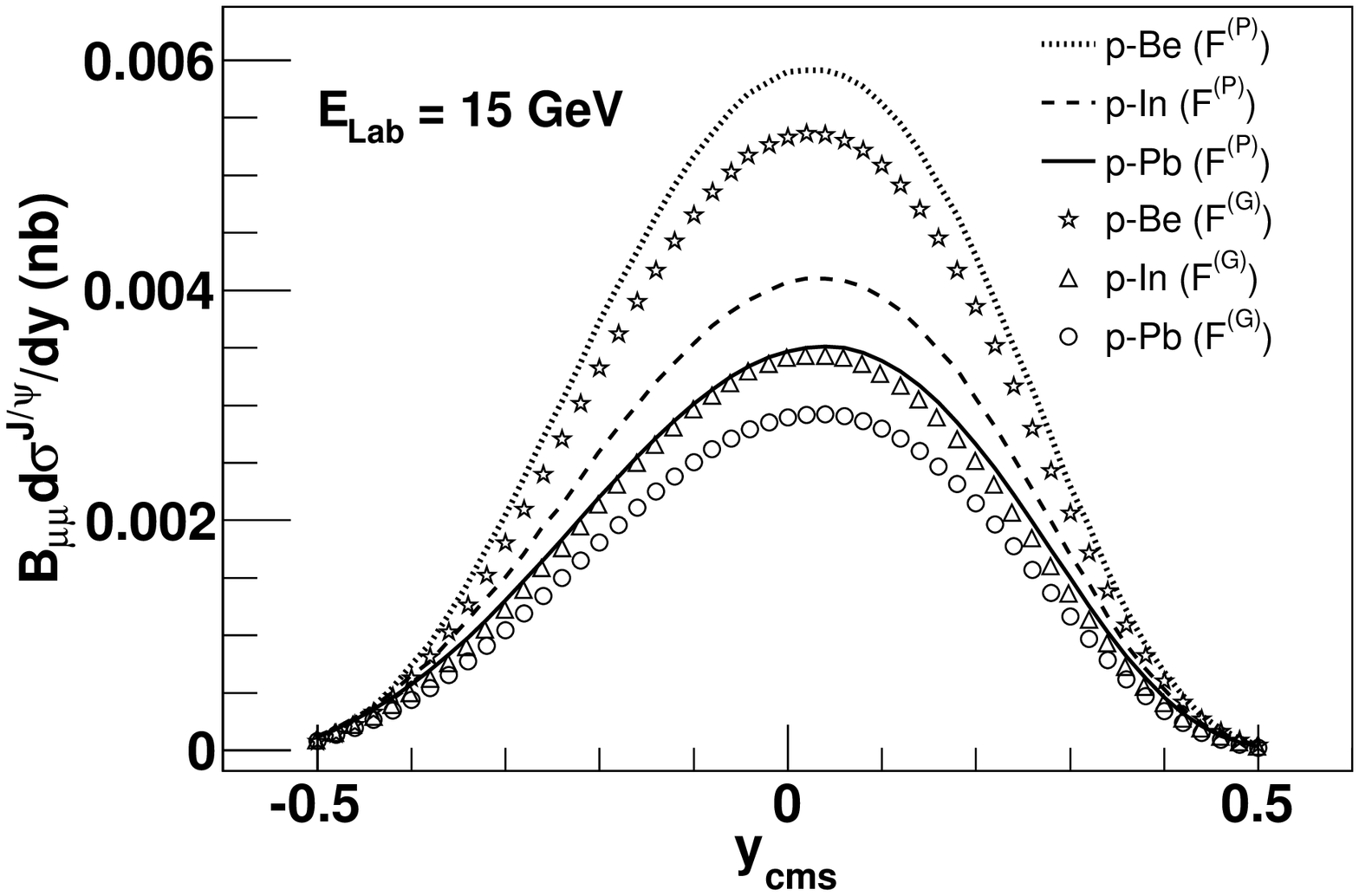}
\includegraphics[width=8.5cm]{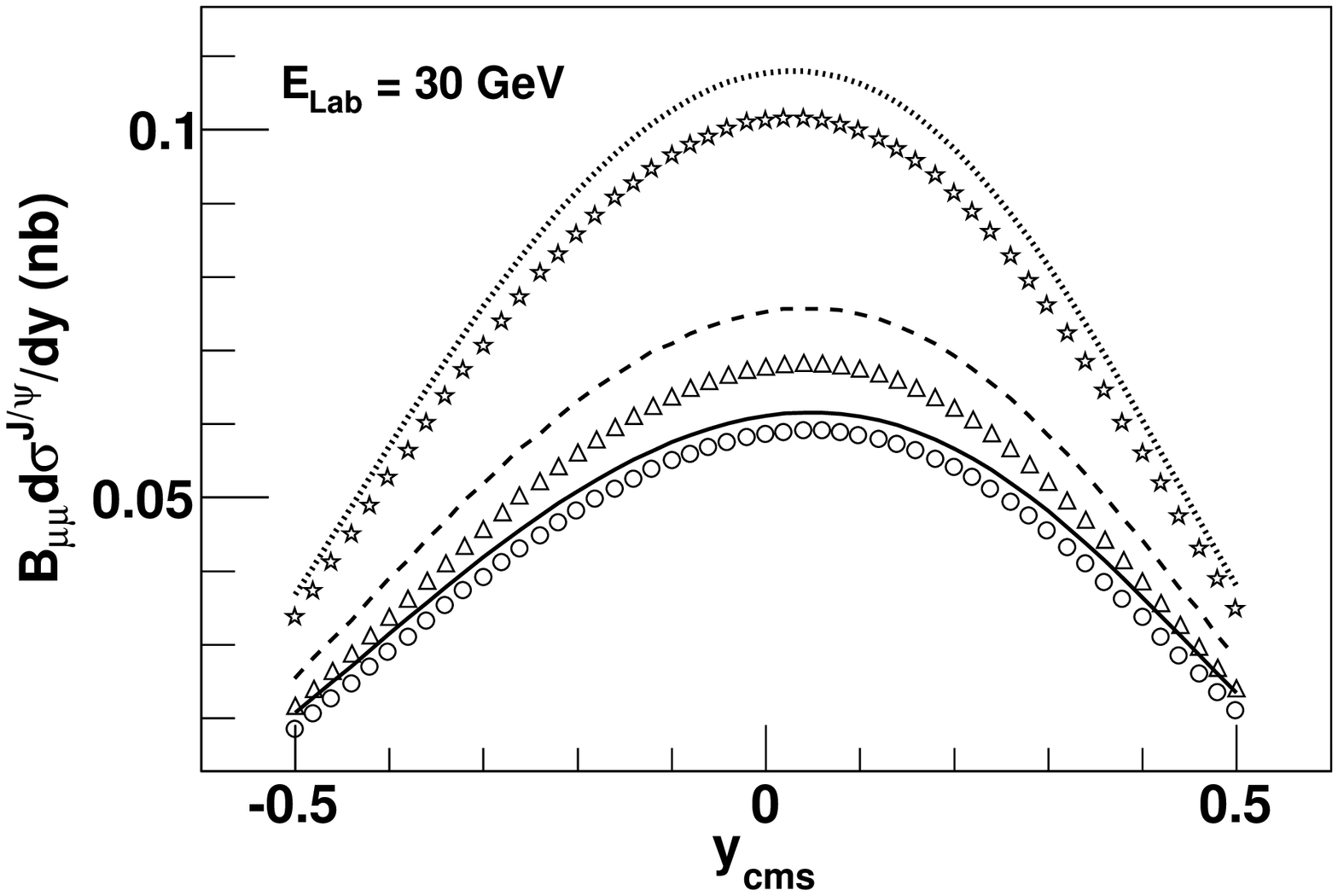}
\includegraphics[width=8.5cm]{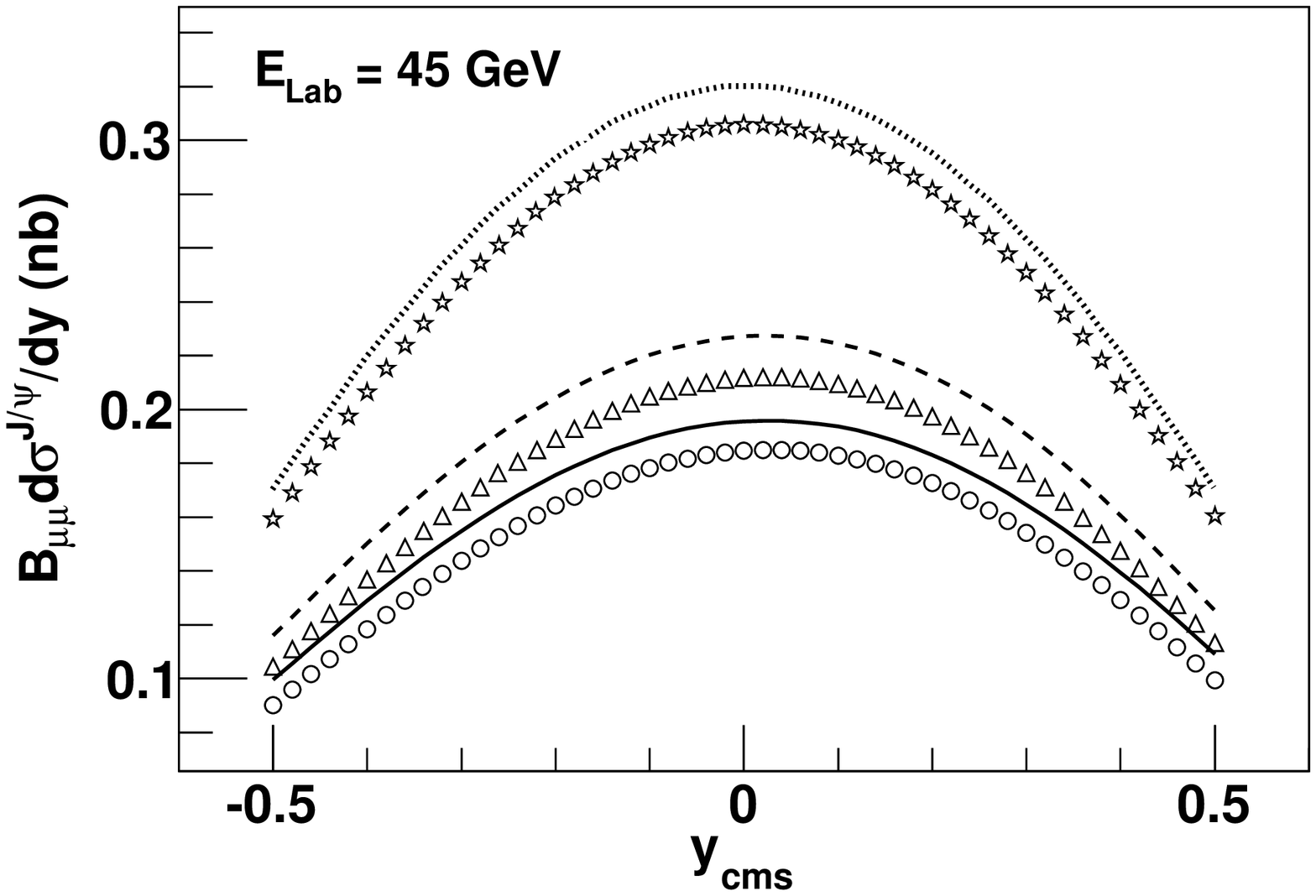}
\caption{\footnotesize The differential J/$\psi$ production cross section per nucleon as a function of the center of mass rapidity $y_{cms}$ for different target nucleus (Be, In and Pb) in the FAIR energy domain, at E$_{Lab}$ = 15 GeV (top), 30 GeV (middle), 45 GeV (bottom), as predicted by our calculations. Both the Gaussian $F^{(G)}(q^2)$  (markers) as well as power law $F^{(P)}(q^2)$  (lines) forms of the transition probabilities are used for calculation of J/$\psi$ yield.}
\label{fig15}
\end{figure*}

Next we present results for charmonium production in p-A collisions at FAIR. For this purpose, we have parameterized the observed dependences of $\varepsilon^2$ on the energy of the incident proton beam for both $F^{(G)}(q^2)$ and $F^{(P)}(q^2)$ using exponential functions. The parametric equations $\varepsilon^2(E_{Lab})$ are shown in Fig.~\ref{fig13}. These parametric relations can now be extrapolated to derive the level of suppression at energies relevant to the CBM experiment at FAIR. In the Fig.~\ref{fig14} we have predicted the normal nuclear suppression in p-A collisions, for $E_{Lab} = $15, 30 and 45 GeV. Following the same approach as NA60, we have calculated J/$\psi$ cross section ratios (with Be as the reference) as a function of $<L>$ for different target nuclei. In case of $F^{(G)}(q^2)$, we can also find the corresponding absorption cross section as might be extracted from the data that will be collected at FAIR. Results are given in Table \ref{tabV}. However the experimentally measured values can be larger than those reported here due to parton shadowing as explained in the last section. In addition to the inclusive production cross sections, we also calculate the J/$\psi$ rapidity distributions for different target nuclei. Fig. \ref{fig15} illustrates the J/$\psi$ differential production cross section at these three energies for three different p-A systems. The obtained distributions can be fitted with the Gaussian functions. The corresponding best fit parameters are given in the Table \ref{tabVIa}.

\begin{table}[ht]\centering
\caption{J/$\psi$ absorption cross section ($\sigma_{abs}$) in cold nuclear matter at FAIR energies. EPS09 parameterization has been used as an input for calculation of J/$\psi$ production and thus the $\sigma_{abs}$ characterizes only final state dissociation.}
\label{tabV}
\vglue4mm
\begin{tabular}{|c|c|} \hline
 E$_{\rm lab}$~[GeV] & $\sigma_{abs}$~[mb] \\ \hline
 15 &  10.4 $\pm$ 1.79 \\ \hline
 30 &  10.1 $\pm$ 1.77  \\ \hline
 45 &  9.87 $\pm$ 1.76   \\ \hline
\end{tabular}
\end{table}

\begin{figure*} \vspace{-0.1truein}
\includegraphics[width=8.5cm]{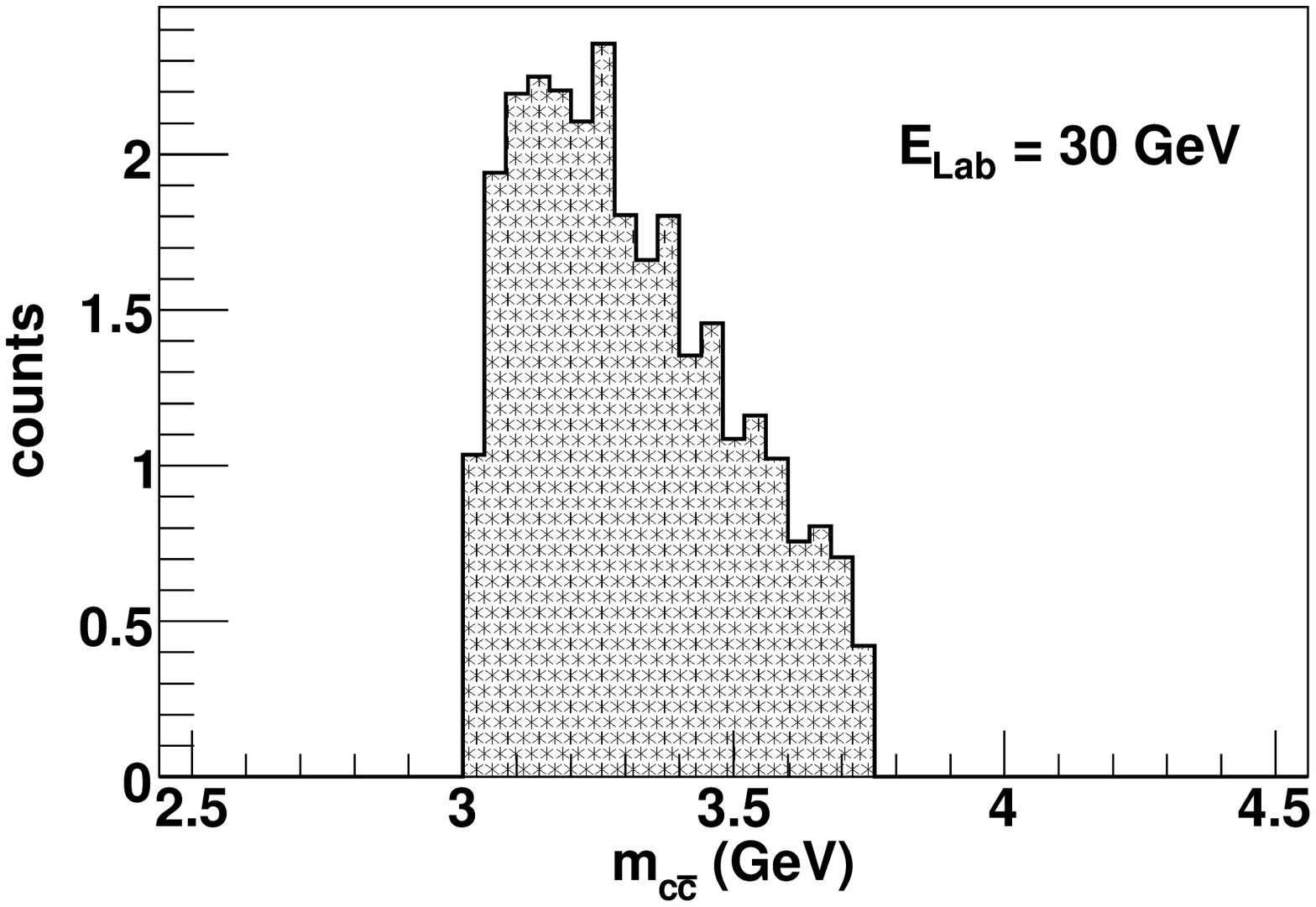}
\includegraphics[width=8.5cm]{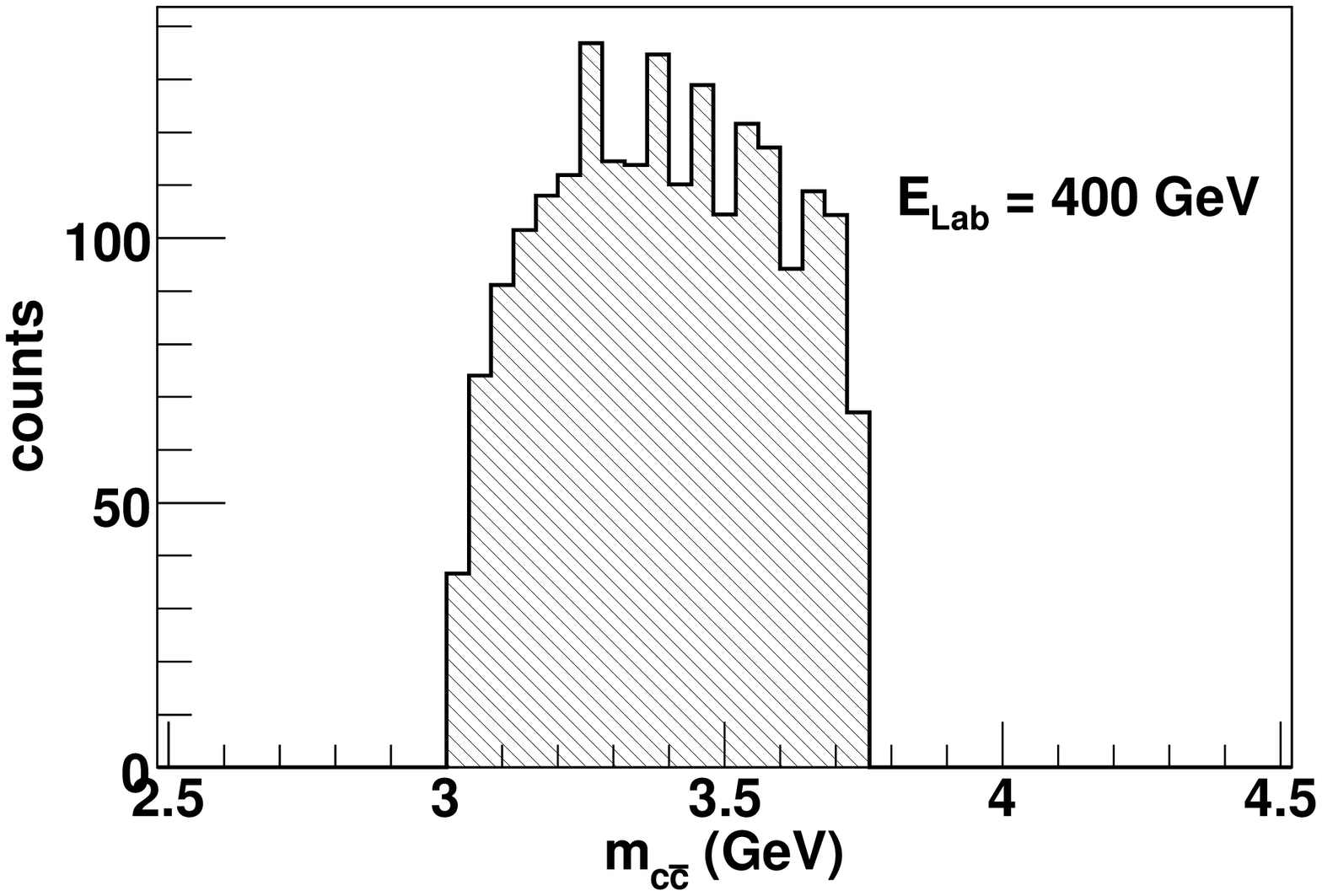}
\caption{\footnotesize The invariant mass distribution of the initially produced $c\bar{c}$ pairs at E$_{Lab}$ = 30 GeV and E$_{Lab}$ = 400 GeV.}
\label{fig16}
\end{figure*}

\begin{figure*} \vspace{-0.1truein}
\includegraphics[width=8.5cm]{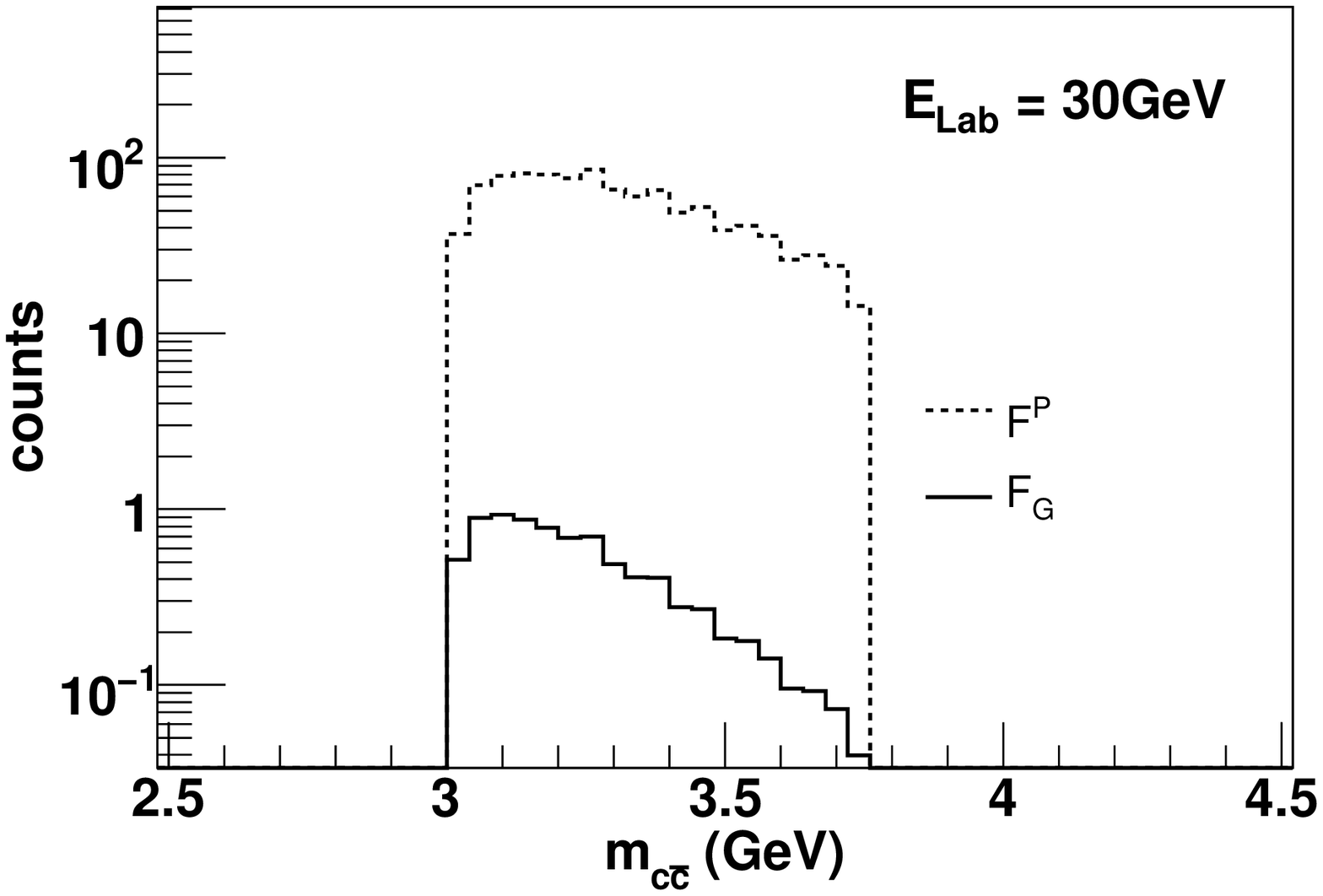}
\includegraphics[width=8.5cm]{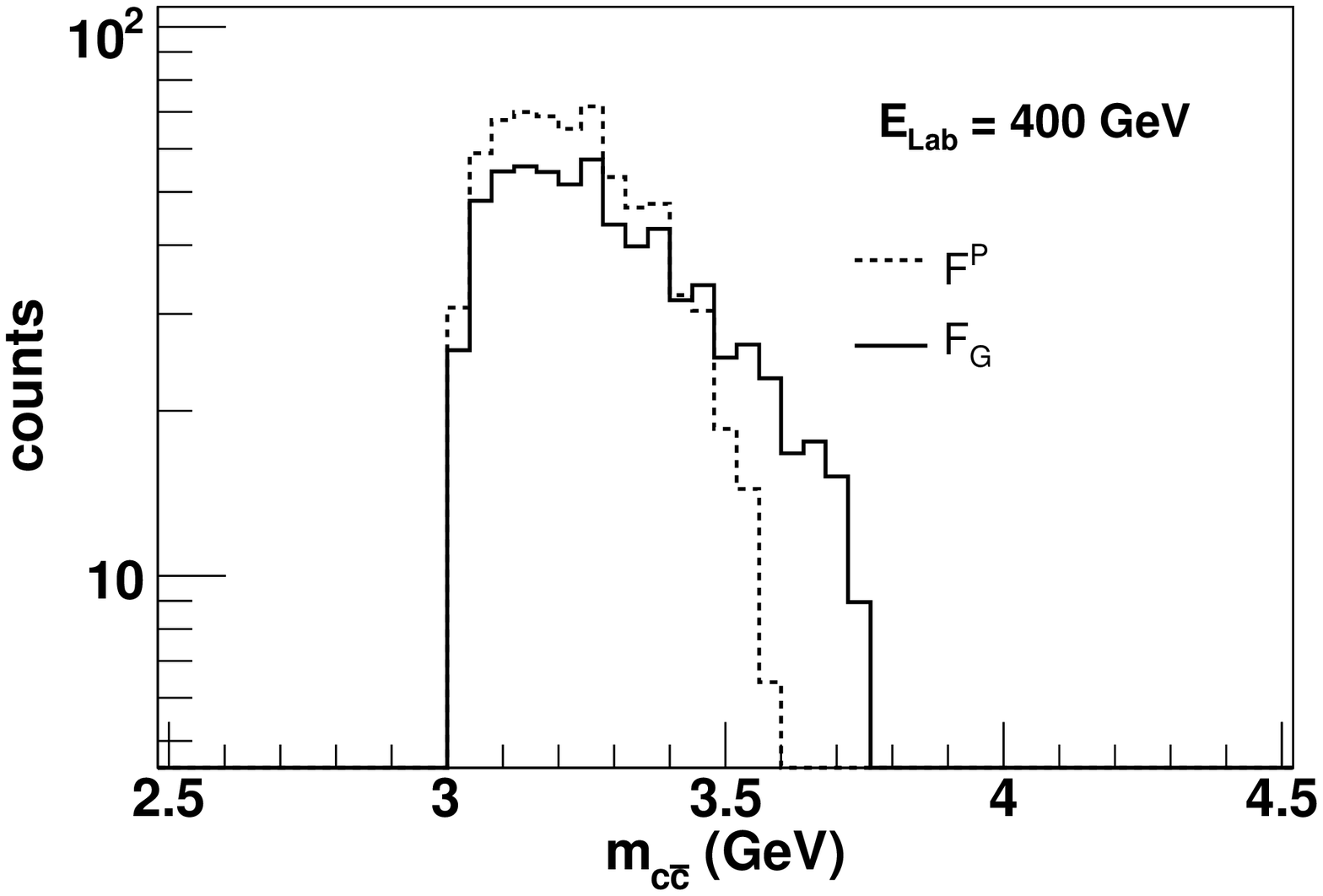}
\caption{\footnotesize The invariant mass distribution of the $c\bar{c}$ pairs at E$_{Lab}$ = 30 GeV and E$_{Lab}$ = 400 GeV, at the final stage after they are convoluted with the transition probabilities $F^{(P)}(q^2)$ and $F^{(G)}(q^2)$ for J/$\psi$ formation.}
\label{fig17}
\end{figure*}

At SPS and higher energies, the calculated inclusive J/$\psi$ yields employing both the forms of transition probabilities, are found to match the measured data for different p-A systems, within error bars. Results for differential distributions are also in close proximity for the two cases. In other none of the inclusive or differential measurements help us to select either of the existing schemes of color neutralization. However the picture appears to be different at lower energies. The two possible hadronization schemes are found to generate different amount of suppressions. The Gaussian distribution gives higher suppression compared to the power law distribution. As the energy of the incident proton beam decreases, the difference between the J/$\psi$ yields as predicted by two models increases. This might be attributed to the difference between the physical mechanisms incorporated in $F^{(G)}(q^2)$ and $F^{(P)}(q^2)$.  The phase space between the thresholds of producing a $c\bar{c}$ pair and open charm meson is fairly large ($\sim$ 5 GeV$^2$). Consequently semi-hard gluons can be radiated during the formation of $c\bar{c}$ pairs with large invariant mass. The radiation reduces the invariant mass of the pair and thus strongly enhances the probability for the pair to form a J/$\psi$ in case of $F^{(P)}(q^2)$. However, for the Gaussian probability $F^{(G)}(q^2)$, the related physical process is not assisted by any gluon radiation during the evolution. As a consequence, the J/$\psi$ formation probability is highly suppressed for the $c\bar{c}$ pairs produced with large invariant mass. This results in a difference between the J/$\psi$ production cross sections following two hadronization schemes, which becomes appreciable at lower beam energies. This can be understood by looking at the invariant mass distribution of the produced  $c\bar{c}$ pairs at different beam energies. Fig.~\ref{fig16} shows the distribution of the $c\bar{c}$ pairs produced at the initial stage (from gluon fusion and light quark annihilation) at the two energies $E_{Lab}$ = 30 GeV and 400 GeV. The shape of the distributions are quite different in the two cases. At 400 GeV the produced pairs are almost uniformly distributed over the allowed phase space interval. However at 30 GeV, the production of large mass $c\bar{c}$ pairs is highly depleted due to the lack of center of mass energy. Each of these input distributions are convoluted with the two different functional forms, $F^{(P)}(q^2)$ and $F^{(G)}(q^2)$, to obtain the final state distributions, as presented in Fig.~\ref{fig17}. At lower energy, the separation between the two distributions, corresponding to two different physical processes of hadronization, becomes higher. This ultimately results in the different production cross sections for two different processes. At FAIR very high statistics J/$\psi$ data will be collected over a fairly large period of time. Hence the measurements are expected to be least suffered by the statistical uncertainties. Thus in addition to the accurate normalization of the heavy-ion data, precise measurements of inclusive as well as differential J/$\psi$ production cross sections in p-p and p-A collisions at FAIR might also help to distinguish the predictions for different scenarios and thereby pinning down the correct mechanism for color-neutralization.

\begin{table*}[ht]\centering
\caption{Parameters of the Gaussian functions used to fit the J/$\psi$ rapidity distributions in p-A collisions for incident proton beam energies 15 GeV, 30 GeV and 45 GeV. The superscripts (P) and (G) respectively denote the power law and Gaussian parameterizations.}
\label{tabVIa}
\vglue4mm
\begin{tabular}{|c|c|c|c|c|} \hline
E$_{Lab}$~[GeV] & $\mu_y^{(P)}$ & $\mu_y^{(G)}$ & $\sigma_y^{(P)}$ & $\sigma_y^{(G)}$\\ \hline
15 & 1.34$\times10^{-2}$ & 1.28$\times10^{-2}$ & 2.11$\times10^{-1}$ & 2.04$\times10^{-1}$ \\ \hline
30 & 1.81$\times10^{-2}$ & 1.82$\times10^{-2}$ & 3.57$\times10^{-1}$ & 3.51$\times10^{-1}$ \\ \hline
45 & 1.78$\times10^{-2}$ & 1.89$\times10^{-2}$ & 4.52$\times10^{-1}$ & 4.41$\times10^{-1}$ \\ \hline
\end{tabular}
\end{table*}

\section{Summary}

The study of charmonium production in high energy collisions is highly advantageous. Because of the large charm quark mass ($m_C >> \Lambda_{QCD}$), and the limitation to be produced at pair (to conserve flavor in strong interaction), charmed hadrons provide us with a new scale to probe QCD. In the present article, we have made a systematic study of J/$\psi$ production in proton induced reactions, in the beam energy range from 158 GeV to 920 GeV, available from different fixed target experiments.The model employed for our study is based on the QCD factorization of inclusive processes. It is general enough to incorporate various existing physical schemes of color neutralization through parametric functions. In case of p-A collisions, two major cold nuclear matter effects namely the modification of parton densities inside the target nucleons at the initial stage and the final state multiple scattering of the nascent $c\bar{c}$ pairs inside the target are taken into consideration. Interplay of these two effects on the measured J/$\psi$ production cross section has been investigated. However a more rigorous theoretical framework should also include other possible nuclear matter effects like parton energy loss at the initial and final stage, that may influence the production. The model gives a satisfactory description of the available data on J/$\psi$ production cross sections in nucleon-nucleon and nucleon-nucleus collisions. However such measurements are not found suitable to discard any of the existing physical mechanisms of color neutralization, when compared to the data. The final state interaction shows a non-trivial dependence on the energy of the incident proton beam. Lower be the energy of the projectile proton, more prominent is the effect of the nuclear medium on J/$\psi$ production. This observation is in agreement with  the existing theoretical and experimental observations and has a direct impact to the heavy-ion collisions. At lower beam energies, the colliding nuclei are less Lorentz contracted and thus results in larger collision time ($t_{coll}$ = $2R_{A}/\gamma$) as well as formation time of the partonic medium. Thus the J/$\psi$ mesons, during its evolution,  are more likely to encounter the (primary) nuclear medium rather than the (secondary) de-confined phase. Results of our present investigations are utilized to make predictions about the level of J/$\psi$ production in p-p and p-A collisions at energies, close to the kinematic threshold. The future FAIR experiments will take data in this energy domain. An accurate measurement of charmonium production production in p-p and p-A collisions performed in the same kinematic conditions of A-A collisions, appears to be an integral part of the charmonium program at FAIR. A proper calibration of the conventional cold nuclear matter effects will help us to identify the unconventional anomalous suppression in nucleus-nucleus collisions that might result from the onset of color de-confinement in the produced system at high baryon densities. A precision measurement of the charmonium production cross sections for different nuclear targets, may also help us to shed light on the much debated issue of color neutralization of a $c\bar{c}$ pair leading to the formation of hidden charm mesons.


\begin{thebibliography}{99}
\bibitem{MS} T. Matsui and H. Satz, Phys. Lett. B{\bf 178}, 416 (1986).

\bibitem{Satz} H. Satz, J. Phys. G32:R25,2006; H. Satz, Rept.Prog.Phys.63:1511,2000

\bibitem{Vogt} R. Vogt, Physics Reports 310, 197 (1999)

\bibitem{NA3} J. Badier {\it et al.}, Z. Phys. C{\bf 20}, 101 (1983).

\bibitem{E772} D. Alde {\it et al.}, Phys. Rev. Lett. {\bf 66}, 133 (1991).

\bibitem{NA38} The NA38 Collaboration, C. Baglin et al., Phys. Lett. B{\bf 220}, 471 (1989); B{\bf 221}, 465 (1990); B{\bf 221}, 472 (1990); B{\bf 225}, 459 (1991).

\bibitem{NA50-200} M.C. Abreu, NA50 Collaboration, Phys. Lett. B{\bf 410}, 337 (1997)

\bibitem{NA50-400} B. Alessandro {\it et al.} NA50 Collaboration, Euro. J.Phys {\bf 48} 329 (2006).


\bibitem{NA50-450} B. Alessandro {\it et al.} NA50 Collaboration, Euro. J.Phys {\bf 33} 31 (2004).

\bibitem{E866} M. J. Leitch {\it et al.} E866 Collaboration, Phys. Rev. Lett. {\bf 84} 3256 (2000).

\bibitem{HERAB} I. Abt {\it et al.} HERA-B Collaboration, Eur.Phys.J.C {\bf 60} 525 (2009) 

\bibitem{NA60} R. Arnaldi {\it et al.} NA60 Collaboration, Phys. Rev. Lett. {\bf 99} 132302 (2007); Roberta Arnaldi, for the NA60 Collaboration, Nucl. Phys. A{\bf 830} 345c-352c, (2009)

\bibitem{CBM} Lect. Notes Phys. {\bf 814}, 1, (2011).

\bibitem{density} I. C. Arsene {\it et al.}, Phys. Rev. C {\bf 75}, 034902 (2007).

\bibitem{Peter} A. Kiseleva, P. P. Bhaduri et. al., Indian J.Phys.85:211-216,(2011), P. Senger, PoS CPOD2009:042,2009.  

\bibitem{Qui} J. Qiu, J.P. Vary and X. Zhang, Nucl. Phys. A698, 571 (2002); Phys. Rev. Lett. {\bf 88} 232301 (2002)

\bibitem{AKC} A. K. Chaudhuri, Phys. Rev. Lett. {\bf 88} 232302 (2002); Phys. Rev. C66 (2002) 021902; Phys. Rev. C68 (2003) 014906; Phys. Rev. C74 (2006) 044907; Nucl.Phys. A734 (2004) 53-56, 


\bibitem{Evap} M.B. Einhorn and S.D. Ellis, Phys. Rev. D{\bf 12}, 2007 (1975);
H. Fritzsch, Phys. Lett. B{\bf 67}. 217 (1977); M. Gl\"{u}ck, J.F. Owens and E. Reya, Phys. Rev. D17, 2324 (1978); J. Babcock, D. Sivers and S. Wolfram, Phys. Rev. D18, 162 (1978).

\bibitem{Octet} G.T. Bodwin, E. Braaten, and G.P. Lepage, Phys. Rev. D{\bf 51}, 1125 (1995).

\bibitem{Singlet} C.-H. Chang, Nucl. Phys. B{\bf 172}, 425 (1980); E.L. Berger and D. Jones, Phys. Rev. D{\bf 23}, 1521 (1981); R. Baier and R. R\"{u}ckl, Phys. Lett. B{\bf 102}, 364 (1981).

\bibitem{CSS} J.C. Collins, D.E. Soper and G. Sterman, Nucl. Phys. B{\bf 263}, 37 (1986).


\bibitem{BQV} C. Benesh, J.Qiu and J.P. Vary, Phys. Rev. C{\bf 50}, 1015 (1994).

\bibitem{cross} H. Fritzsch, Phys. Lett. B{\bf 67}, 215 (1977); M. Gluck, J. F. Owens , and E. Reya, Phys. Rev. D{\bf 17}, 2324 (1978).


\bibitem{MSTW} A.D. Martin,W.J. Stirling, R.S. Thorne, G. Watt, Eur.Phys.J.C63:189-285,2009; A.D. Martin, W.J. Stirling, R.S. Thorne, G. Watt, Eur.Phys.J.C64:653-680,2009; A.D. Martin, W.J. Stirling, R.S. Thorne, G. Watt, Eur.Phys.J.C70:51-72,2010.

\bibitem{EPS09} K.J. Eskola, H. Paukkunen and C.A. Salgado, JHEP04 (2009) 065

\bibitem{Lourenco} C. Lourenco, R. Vogt and H. K. Wohri, JHEP {\bf 0902} 014 (2009)

\bibitem{QVZ} J.-W. Qiu, J.P. Vary and X.-F. Zhang, in preparation.

\bibitem{LQS} M. Luo, J.-W. Qiu and G. Sterman, Phys. Rev. D{\bf 49}, 4493 (1994).

\bibitem{LABref} D. Kharzeev, C. Lourence, M. Nardi and H. Satz, Z. Phys. C{\bf 74}, 307 (1997).

\bibitem{EKS98} K.J. Eskola, V. J. Kolhinen and C.A. Salgado, Eur.Phys.JC{\bf 9}61 (1999)

\bibitem{EPS08} K.J. Eskola, H. Paukkunen and C.A. Salgado, JHEP0807 (2008) 102

\end{thebibliography}
\end{document}